\documentclass[nonatbib]{elsarticle}
\usepackage[utf8]{inputenc}
\usepackage[british]{babel}
\usepackage{csquotes}
\usepackage{subfiles}
\usepackage{booktabs}
\usepackage{graphicx}
\usepackage{multirow}
\usepackage{lscape}
\usepackage{setspace}
\usepackage{pdflscape}   
\usepackage{lscape}
\usepackage{eurosym}
\usepackage[labelfont=bf, font=footnotesize]{caption}
\DeclareCaptionLabelSeparator{vline}{$~\vert~$}
\usepackage[table,xcdraw]{xcolor}
\usepackage[hidelinks]{hyperref}
\hypersetup{colorlinks=true,citecolor={blue},citebordercolor={blue}}
\usepackage{geometry}
\makeatletter
\let\c@author\relax
\makeatother
\usepackage[backend=biber, style=numeric-comp, autocite=superscript, maxcitenames=1, isbn=false, date=year, url=false, sorting=none]{biblatex}
\DeclareAutoCiteCommand{superscript}{\supercite}{\supercites}
\AtEveryBibitem{\clearfield{note} \clearlist{language}}
\makeatletter
\def\ps@pprintTitle{%
  \let\@oddhead\@empty
  \let\@evenhead\@empty
  \let\@oddfoot\@empty
  \let\@evenfoot\@oddfoot}
\makeatother

\title{Power sector models featuring individual BEV profiles: \\ Assessing the time-accuracy trade-off\tnoteref{t1}}

\author[1,2,3]{Adeline Guéret\corref{cor1}}

\cortext[cor1]{Corresponding author: \texttt{agueret@diw.de}.}
\affiliation[1]{organization={DIW Berlin},
addressline={Anton-Wilhelm-Amo-Straße 58},
city={10117 Berlin},
country={Germany}}

\affiliation[2]{organization={Technische Universität Berlin},
addressline={Straße des 17.\ Juni 135},
city={10623 Berlin},
country={Germany}}

\affiliation[3]{organization={OFCE - Sciences Po Paris},
addressline={10, Place de Catalogne},
city={75014 Paris},
country={France}}

\begin{document}


\begin{abstract}
Electrifying passenger cars will impact future power systems. To understand the challenges and opportunities that arise, it is necessary to reflect ``sector coupling'' in the modeling space. This paper focuses on a specific modeling approach that includes dozens of individual BEV profiles rather than one aggregated BEV profile. Although including additional BEV profiles increases model complexity and runtime, it avoids losing information in the aggregation process. We investigate how many profiles are needed to ensure the accuracy of the results and the extent to which fewer profiles can be traded for runtime efficiency gains. We also examine whether selecting specific profiles influences optimal results. We demonstrate that including too few profiles may result in distorted optimal solutions. However, beyond a certain threshold, adding more profiles does not significantly enhance the robustness of the results. More generally, for fleets of 5 to 20 million BEVs, we derive a rule of thumb consisting in including enough profiles such that each profile represents 200,000 to 250,000 vehicles, ensuring accurate results without excessive runtime.
\end{abstract}

\doublespacing

\maketitle
\renewcommand{\thefootnote}{\alph{footnote}}

\section{Introduction}

Battery electric vehicles (BEVs) have significant potential to mitigate climate change, as they offer an efficient way to reduce greenhouse gas emissions in the passenger road transport sector\autocite{vsimaitis2025battery, schreyer2024distinct, sacchi2022and, rottoli2021alternative}. Electrifying the light-duty vehicle fleet will increasingly intertwine the previously separate transport and electricity sectors. This evolution toward ``sector coupling'' must be reflected in modeling if we want to better understand and anticipate the challenges and opportunities ahead\autocite{muratori2020future, richardson2013electric, galus2012integrating}. 

From a modeling perspective, integrating BEVs and power systems remains a complex endeavor, particularly since spatiotemporal scales can differ significantly. Agent-based transport models focus on understanding travel behavior at a microscopic scale\autocite{kagho2020agent}. These simulations can typically generate travel activities with a resolution of one minute for exact geographic coordinates and a very large number of agents. Travel diaries and more recent related statistical methods also leverage very precise data, e.g., exact times and locations of trips, in particular since developments in big data techniques have opened the door to GPS tracking and mobile phone data collection\autocite{anda2017transport}. This level of granularity is well-suited for analyzing the impacts of BEVs on local or regional distribution grids\autocite{li2024impact, reibsch2024low, steinbach2024grid, waraich2013plug}. However, for other applications such as coupling with capacity expansion and/or generation power sector models, this level of granularity is too high, as these models are usually defined at an hourly resolution or lower and rely on aggregated data for local regions or entire countries\autocite{prina2020classification}. Including a representation of BEVs in capacity expansion or dispatch models therefore requires abstraction of some details, in the form of data aggregation or of stylized assumptions regarding trip patterns, plug-in or charging behaviors, etc. This has led to the development of stochastic tools to generate synthetic BEV mobility data. Inspired by agent-based transport models, these tools usually harvest empirical data from travel surveys and create additional modules to compute the data required by power sector models, such as driving electricity consumption and power rating time series\autocite{mahmud2016review}.

The first attempts to integrate electric mobility in power sector models date back to the late 2000s\autocite{lund2008integration, mathiesen2008integrated}. These attempts included limited detail regarding the transport modeling side. For example, individual car driving profiles were not considered. Instead, an overall transport demand profile and one single large battery were modeled. Additionally, assumptions were made regarding the proportion of the fleet that is driving at certain times of the day and the proportion of parked vehicles that are effectively connected to the grid. Later developments increasingly acknowledged the importance of taking into account car travel behaviors and induced vehicle charging patterns. This allows for a more accurate representation of BEVs' availability and load, enabling a better understanding of conflicts and synergies with other time-varying factors, such as variable renewable energy sources (vRES)\autocite{powell2022charging, wolinetz2018simulating, schauble2017generating, schill2015power, harris2012temporal, kiviluoma2011methodology}. 

Although representing a diversity of BEV profiles in power sector models is methodologically established by now, there are still different approaches to integrating them more precisely. One approach is to aggregate a large number of very detailed mobility profiles. The resulting aggregated time series of, e.g., driving or charging electricity consumption, and, if applicable, grid availability, and battery storage level are then included in power sector models as if there were only one representative vehicle\autocite{frank2025potential, syla2024optimal, mangipinto2022impact, brown2018synergies}. This approach encompasses a great wealth of driving profiles while maintaining relatively low additional model complexity. However, the aggregation step results in the loss of some information contained in individual profiles. This has led to the development of alternative aggregation methods\autocite{muessel2023accurate}. Another approach is to include several (dozens of) individual BEV profiles scaled to match a given fleet size\autocite{gueret2025moderate, gueret2024impacts, gaete2024power}. With this approach, accuracy at the individual profile level is not lost in aggregation because each profile retains its internal coherence across mobility patterns, driving electricity consumption, grid availability, and battery storage level dynamics. However, including additional individual BEV profiles is computationally demanding, and potentially intractable, especially if representativeness at the fleet level requires including a very large number of profiles. 

In this work, we focus on the second approach, the ``individual BEV profiles'' approach. We do not intend to compare how different approaches behave. Using an open-source stochastic tool to generate synthetic BEV profiles and an open-source power sector model, we investigate how increasing the number of BEV profiles included in a power sector model affects runtime and the accuracy of optimal solutions. Additionally, we analyze the role of profile selection, i.e., whether and to what extent results change when drawing alternative BEV profile samples, for a given number of profiles. 

We show that, beyond 60-80 BEV profiles, runtime increases steeply with the overall number of profiles. However, our findings underscore the importance of including at least 20 BEV profiles to avoid overestimating the cost impact of BEVs when modeling fleets with 5 to 20 million BEVs. Additionally, we find that optimal capacity mixes vary significantly with the number of profiles included. Our results suggest that ensuring robust optimal capacity results requires including more BEV profiles than are needed for stable cost results, particularly for stationary Li-ion battery technologies. Provided that enough profiles are included, profile selection only marginally influences optimal results for most technologies, except for solar photovoltaic (PV), for which optimal capacity depends more on profile selection when BEVs charge smartly. We derive a rule of thumb according to which including enough profiles such that each BEV profile represents between 200,000 and 250,000 BEVs ensures reasonable runtimes and accurate optimal results for fleets of 5 to 20 million BEVs in various settings.  

\section{Results}
\label{sec:results}

\subsection{Scenarios}

In our study, scenarios refer to a system including BEVs, while the reference represents a system without BEVs. Nothing else except the introduction of BEV profiles differs between the reference and scenarios. How BEV profiles are generated and what they exactly represent is explained in more details in Section~\ref{sec:methods}. Main results reflect a 2030 setting for Germany as an island with no modeling of sector coupling technologies such as heat pumps or green hydrogen, except for a 15 million fleet of electric vehicles in scenarios. Two alternative settings are also analyzed in order to put in perspective main results: in one setting, an industrial demand for green hydrogen is included; in another, power sectors of neighbouring countries and interconnections are introduced. Section~\ref{sec:methods} provides further information relative to all considered settings.

We design scenarios that include an increasing quantity of BEV profiles, for a constant BEV fleet of 15 million vehicles. Amounts of profiles considered range from 5 to 120 at an increment of 5. Additionally, for $x$ a given number of BEV profiles among the ones considered, we randomly draw ten samples of $x$ BEV profiles from a fixed pool of 200 BEV profiles. We consider two charging strategies, namely smart charging and bidirectional charging. BEV profile samples used for a given number of BEV profiles are the same irrespective of the charging strategy in order to ensure comparability across them. Hence, for a given charging strategy in a given setting, we analyze 240 scenarios. We carry out additional sensitivity analyses with BEV fleets of 5, 10 and 20 million vehicles.

\vspace{0.25cm}
\subsection{Increasing the number of included BEV profiles noticeably increases total runtimes}
\vspace{0.25cm}

As each additional BEV profile implies additional variables and constraints, increasing the number of BEV profiles leads to larger models that are more complex to solve. Since total runtime is a decisive factor in modelling, we first measure how much time it takes to solve problems including more or less BEV profiles. In order to ensure comparability across results, we consider the total runtime that is needed from data loading to model computation and solving. Besides, in order to avoid any variation in computation performance due to ``warm-up'' or unequal memory use, we make sure to start each time from scratch. 

\begin{figure}[!ht]
    \centering
    \includegraphics[width=1\linewidth]{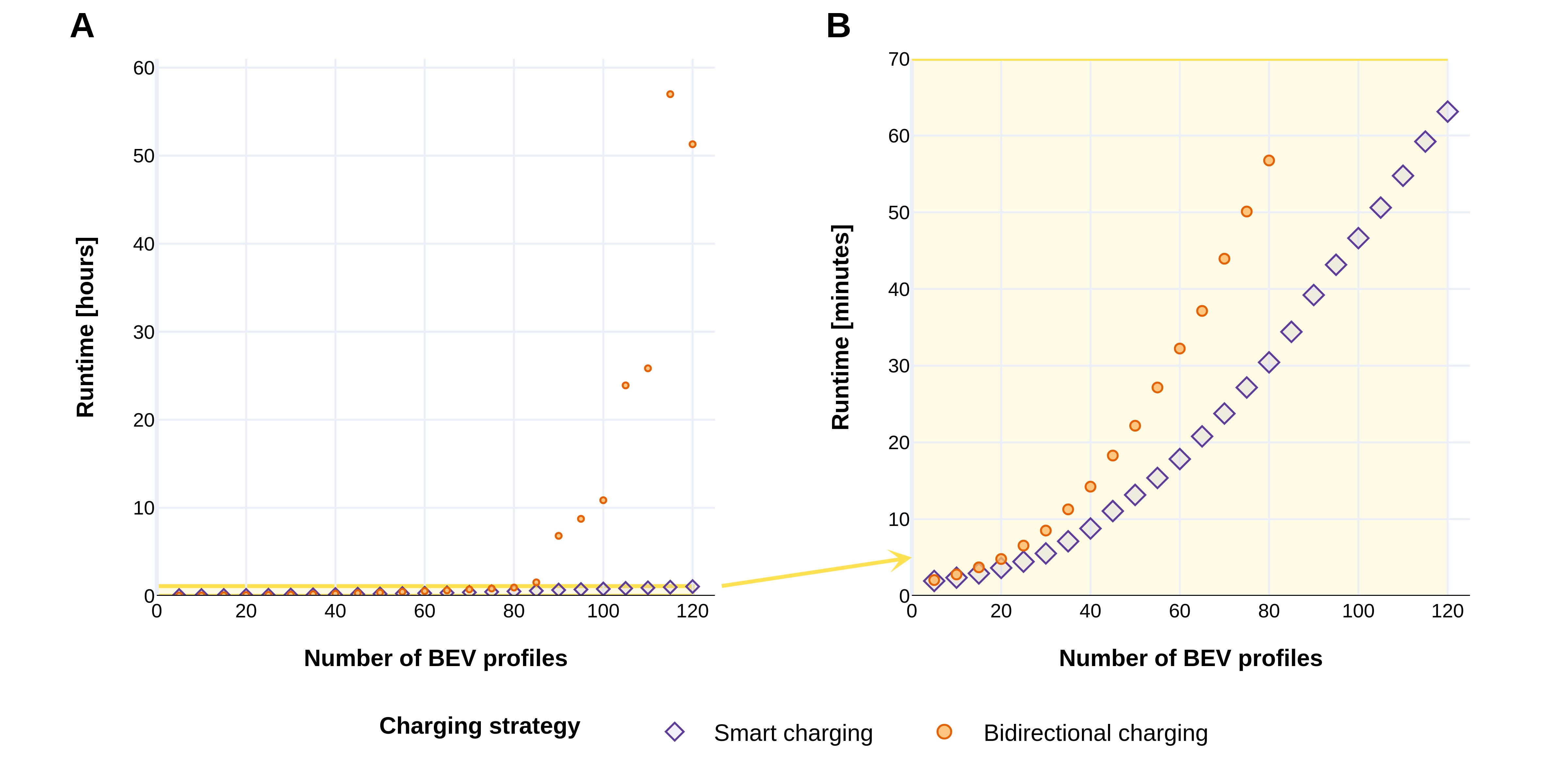}
    \caption{
    \textbf{Runtime by number of BEV profiles and charging strategy [Main setting].} 
    Runtime encompasses all steps from loading data to model computation and solving. Runtime in minutes for a selection of scenarios in the main setting (\textbf{A}). Zoom on runtimes up to 70 minutes (\textbf{B}). Scenarios consider Germany as an island, a fleet of 15 million BEVs and no industrial green hydrogen demand.}
    \label{fig:runtimes_base}
\end{figure}

As shown in Figure~\ref{fig:runtimes_base} for a setting with no interconnections nor green hydrogen demand, total runtime increases with the number of BEV profiles included in the model for both charging strategies. While the runtime increase remains moderate with smart charging, it is particularly strong for bidirectional charging when more than 80 BEV profiles are included. Up to 80 profiles, the runtime remains below an hour but is almost 10 times larger (about 10 hours) with 100 BEV profiles and 50 times larger with 120 profiles.

For a given amount of BEV profiles, alternative profile selection does not significantly impact runtime (Figure~\ref{fig:runtimes_samples}). The more profiles are included, the more runtime differs across samples for a given number of profiles. However, profile selection does not explain jumps observed in Figure~\ref{fig:runtimes_base} between, e.g., 100 and 105 profiles or 110 and 115 profiles. Rather, these jumps are likely driven by other model constraints such as capacity bounds. This is confirmed by the fact that such jumps are less striking in a very stylized setting with no capacity nor energy constraints and only solar photovoltaic, wind onshore, wind offshore and gas turbines (OCGT and CCGT) as available technologies (Figure~\ref{fig:runtimes_stylized_othersettings}, panels A and B). Nevertheless, even in a more stylized setting, runtime increases strongly with the number of BEV profiles, with large increases when including more than 80 profiles. Similarly, this finding holds for other settings with more flexibility options such as an industrial demand for green hydrogen or interconnections with neighbouring countries (Figure~\ref{fig:runtimes_stylized_othersettings}, panels C and D). Interestingly, in an interconnected setting, runtime increases significantly not only for bidirectional charging but also for smart charging, starting already at 60 profiles (respectively 70 profiles for smart charging). 

Beyond a threshold of 60 or 80 profiles depending on the setting considered, the increase in time needed to solve the model with additional BEV profiles becomes considerable. Whether it is worth incurring these additional time costs depends on the extent to which including more BEV profiles improves the overall results accuracy.

\vspace{0.25cm}
\subsection{Including too few profiles overestimates cost impacts and underestimates cost savings brought by BEVs}
\vspace{0.25cm}

\begin{figure}[!ht]
    \centering
    \includegraphics[width=1\linewidth]{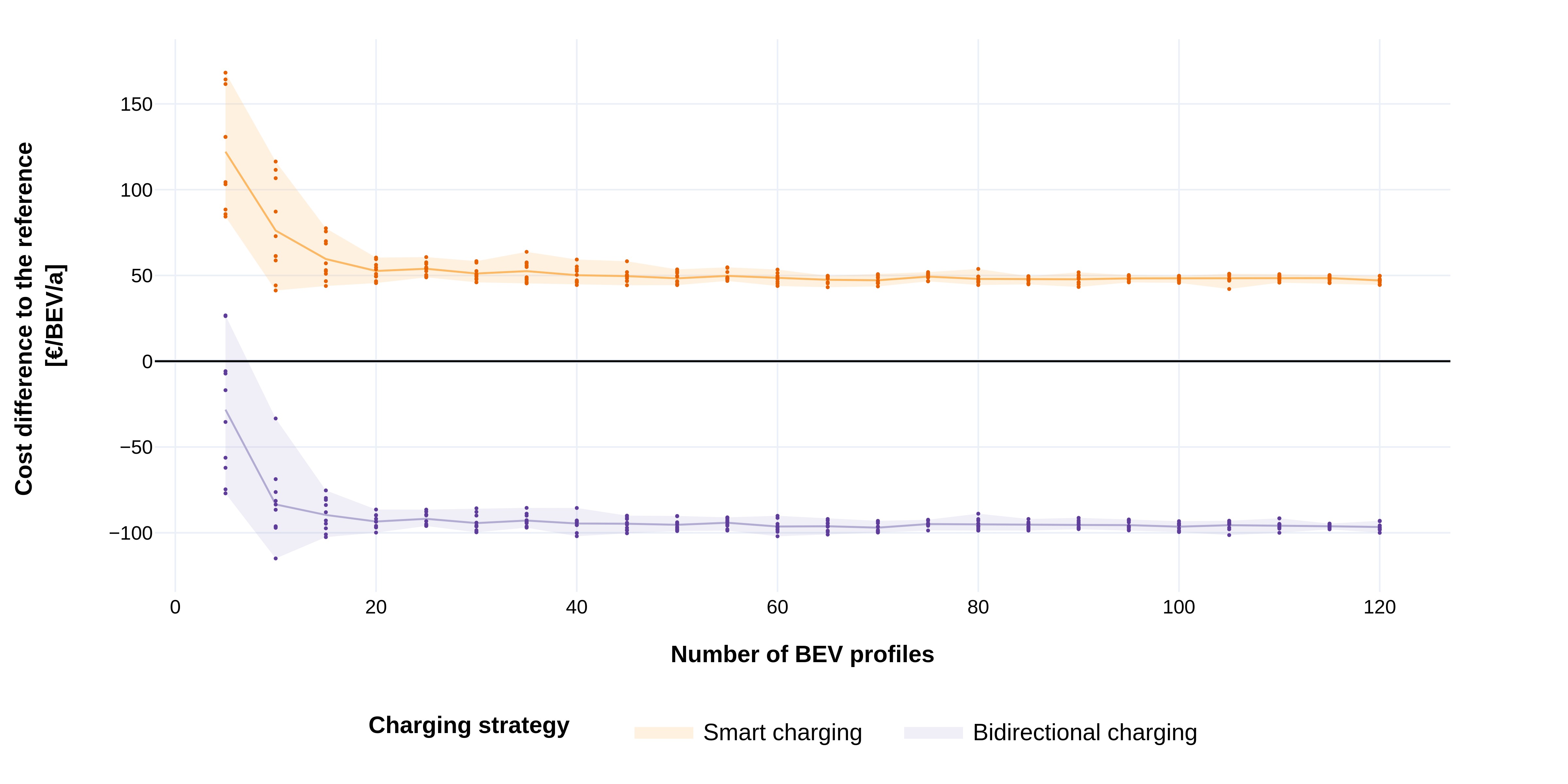}
    \caption{
    \textbf{System cost difference to the reference by number of BEV profiles and charging strategy [Main setting].}
    Differences are expressed relative to the reference with no BEVs. For $x$ a given number of BEV profiles, each dot refers to the optimal solution of a model including a randomly drawn sample of $x$ BEV profiles. Lines connect sample averages for a given charging strategy. Shaded areas delimit the empirical cost difference interval for a given charging strategy.
    }
    \label{fig:systemcosts_both}
\end{figure}

For both charging strategies, the number of BEV profiles included influences the optimal system costs. Cost differences to the reference, averaged over scenarios including the same amount of profiles and the same charging strategy, depend on the amount of BEV profiles included (Figure~\ref{fig:systemcosts_both}, solid lines). With 20~profiles up to 120~profiles, average cost differences appear to be robust to the number of BEV profiles included and stabilize at additional costs of around 50~euros per BEV per year for smart charging and at 100~euros of cost savings per BEV per year for bidirectional charging. From ~5 up to 20~profiles, the smaller the number of profiles included, the larger the average cost difference to the reference. For smart charging, this means that including less than 20~BEV profiles leads to overestimate the cost impacts of BEVs. The average cost difference to the reference is about 2.5~times bigger when including 5~BEV profiles compared to when including 20~profiles or more. Similarly, for bidirectional charging, including less than 20~BEV profiles leads to underestimate the cost savings brought by electric vehicles: with 5~profiles, the average cost savings (negative costs) are about four times smaller than with at least 20~profiles. These results also hold in other settings including an industrial green hydrogen demand or explicitly modeling neighbouring countries' interconnected power sectors (Figure~\ref{fig:systemcosts_settings}).

In samples with a small number of profiles, individual profile characteristics are scaled by a larger factor since we model a fleet of constant size of 15 million BEVs. This means that each profile has more weight in shaping a sample's aggregate BEV parameters such as the aggregate BEV load or BEV battery capacity, which can lead to a greater diversity of samples when it comes to aggregate BEV characteristics (Figure~\ref{fig:bev_delta_aggregates}). Nevertheless, we find that previously detailed results still hold, even when excluding all samples deviating from mean values by more than~5\% (Figure ~\ref{fig:systemcosts_trimmed_both}).

Looking beyond the sample average, cost differences to the reference depend more on profile selection when the number of profiles is small, i.e., less than~20. In the case of bidirectional charging, this can even lead to qualitatively different results. Indeed, for one sample with 5~profiles, cost differences are even positive. Interestingly, when including only 5~profiles, despite a larger variability across samples, the smallest cost difference to the reference remains larger than the highest outcomes of samples including at least 20~profiles. This is however not the case when including 10 or 15 profiles, in which cases some samples even show smaller cost impacts than in cases with more profiles. With at least 20~profiles, profile selection plays a much smaller role as optimal outcomes vary much less across samples for a given amount of profiles. 

\vspace{0.25cm}
\subsection{Optimal stationary Li-ion battery and natural gas capacity critically depend on the number of profiles}
\vspace{0.25cm}

\begin{figure}[!ht]
    \centering
    \includegraphics[width=1\linewidth]{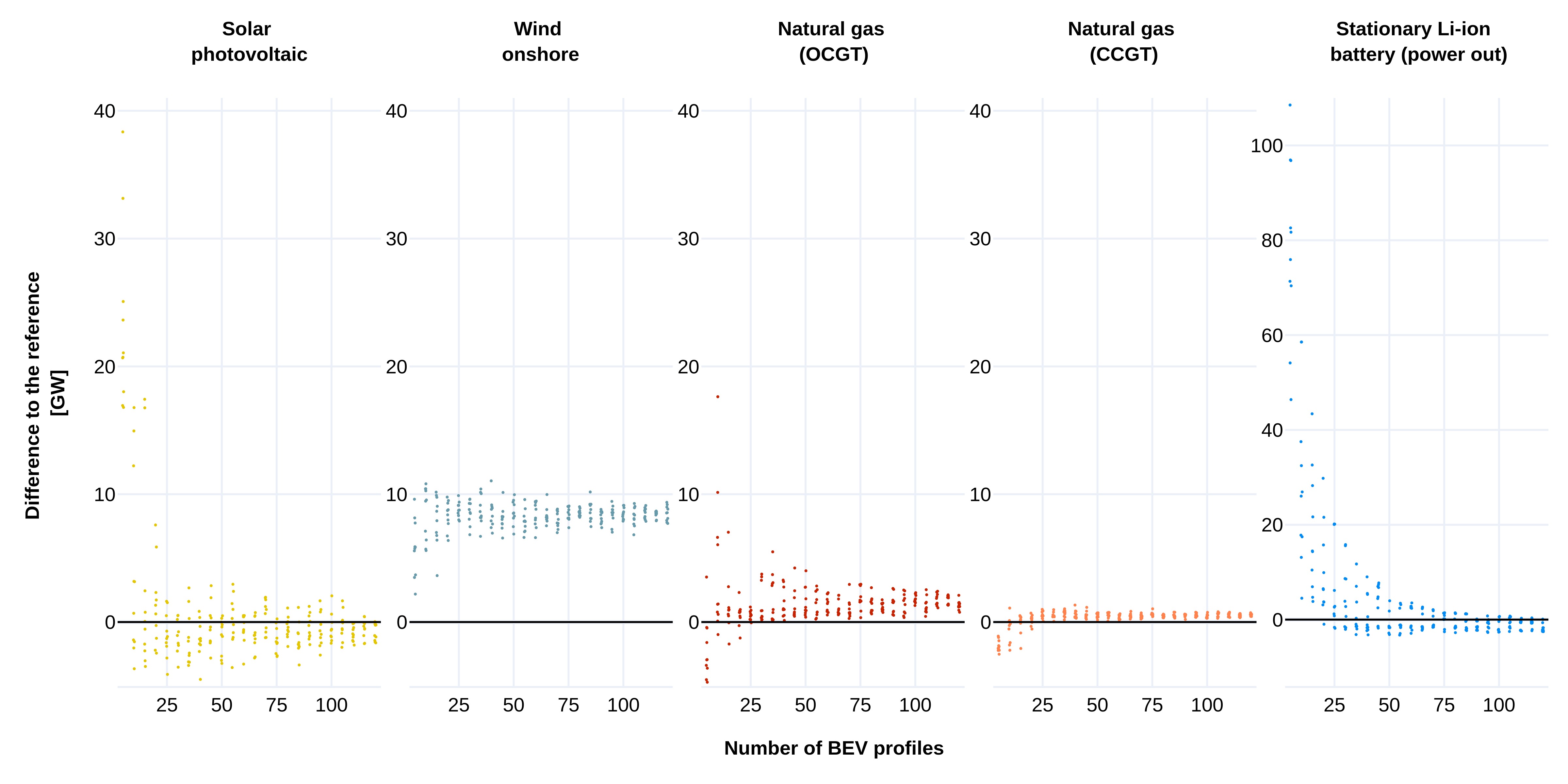}
    \caption{
    \textbf{Change in optimal capacity for selected technologies when BEVs charge smartly [Main setting].}
    Changes are expressed relative to the reference with no BEVs. For $x$ a given number of BEV profiles, each dot refers to the optimal solution of a model including a randomly drawn sample of $x$ BEV profiles. 
    }
    \label{fig:capacity_smart}
\end{figure}

\begin{figure}[!ht]
    \centering
    \includegraphics[width=1\linewidth]{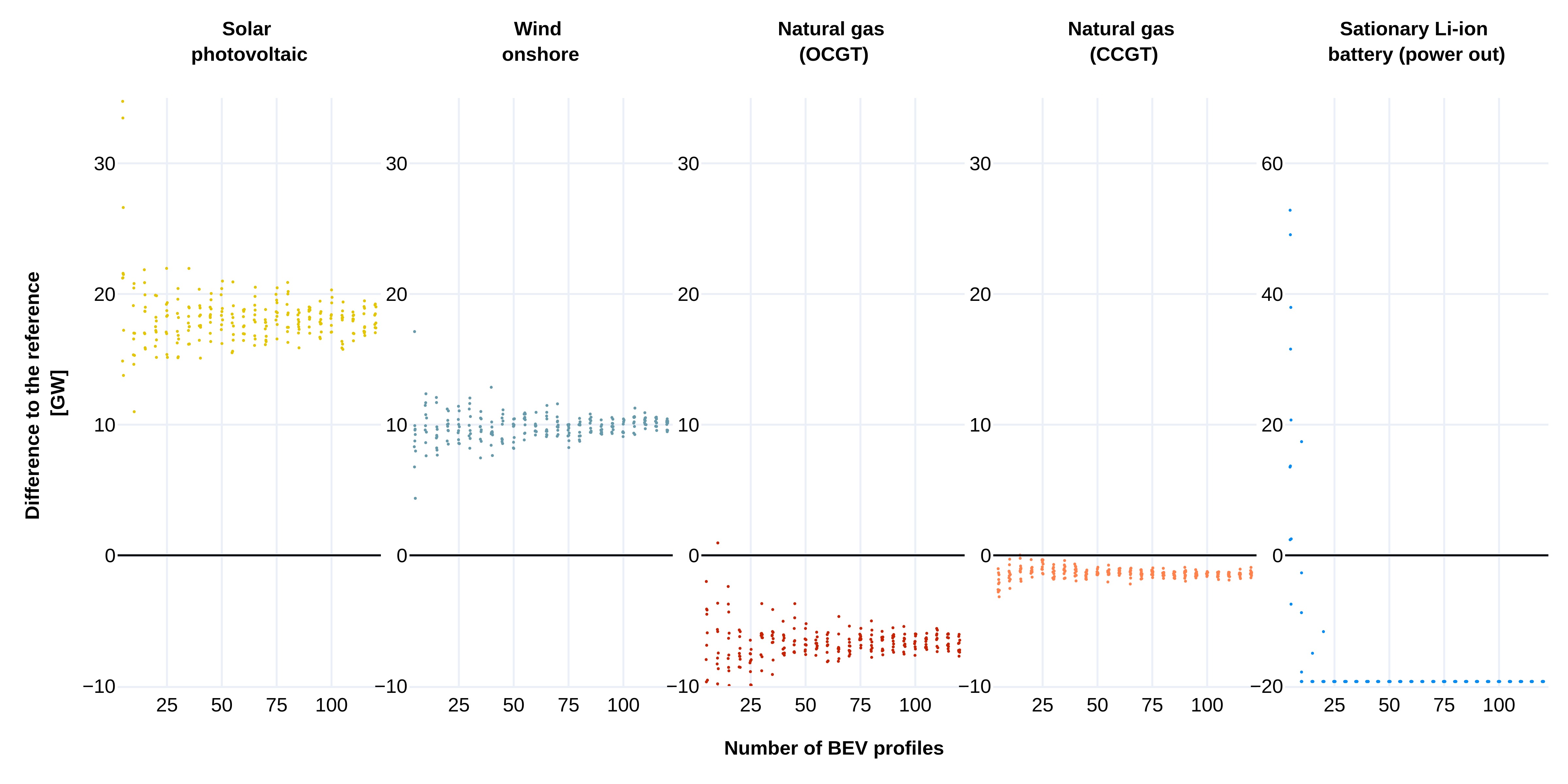}
    \caption{
    \textbf{Change in optimal capacity for selected technologies when BEVs charge bidirectionally [Main setting].}
    Changes are expressed relative to the reference with no BEVs. For $x$ a given number of BEV profiles, each dot refers to the optimal solution of a model including a randomly drawn sample of $x$ BEV profiles. 
    }
    \label{fig:capacity_v2g}
\end{figure}

For a given number of BEV profiles, selecting different BEV profiles leads to different optimal capacity mixes. Nevertheless, the extent of the variation across samples of similar sizes depend on the number of BEV profiles, the technology and the charging strategy considered (Figure~\ref{fig:capacity_smart} and Figure~\ref{fig:capacity_v2g}). 

For both charging strategies, the optimal capacity for stationary Li-ion battery show that including too few BEV profiles can lead to conclusions that contradict those drawn when including more profiles. When a low amount of BEV profiles is included, the integration of BEVs in the power sector leads to a strong increase in the optimal power outflow capacity of stationary Li-ion battery in all or almost all samples (Figure~\ref{fig:capacity_smart} and Figure~\ref{fig:capacity_v2g}, utmost right panel) compared to the reference (Figure~\ref{fig:capacity_reference}). This is particularly striking for cases with 5 profiles, where large additional Li-ion battery power capacity undoubtedly contributes to the higher system costs differences detailed previously. For smart charging, even for cases with up to 30 BEV profiles, almost all samples show a rather large increase in optimal stationary Li-ion battery power outflow capacity. In contrast, with 50 BEV profiles or more, changes in the optimal Li-ion battery power capacity are much smaller in absolute terms and equally split between positive and negative values. With more than 90 profiles, almost no sample shows a capacity increase. 

For bidirectional charging, Li-ion battery power capacity decreases compared to the reference in all samples when 15 profiles or more are included and the technology totally disappears from the mix in all samples with at least 25 profiles. Hence, with too few profiles, power capacity effects for Li-ion battery are not only overestimated but may also qualitatively diverge from findings derived from models with a more detailed BEV representation. Optimal Li-ion battery energy capacity (Figure~\ref{fig:capacity_battery}) and dispatch (Figure~\ref{fig:dispatch_battery}) show similar patterns. 

Similar effects are at play for the optimal generation capacity of combined cycle gas turbines (CCGT), in particular for smart charging. In this case, while a small amount of profiles leads to a decrease in optimal capacity with respect to the reference, including more profiles ($\geq 25$) constantly yields a slight capacity increase for this technology. Changes in the optimal aggregate CCGT electricity generation are consistent with capacity effects (Figure~\ref{fig:dispatch_reference} and Figure~\ref{fig:dispatch_generation}). The optimal capacity of open cycle gas turbines (OCGT) is also sensitive to the number of profiles included and results accuracy also benefits from including at least 25 profiles. 

Overall, ensuring accurate optimal capacity results requires to include at least 25 profiles, more than what is required for robustness in cost results. One noticeable exception is stationary Li-ion battery with smart charging, for which including at least 75 profiles proves useful. 

For both charging strategies, including less BEV profiles lead to a greater range in optimal capacity changes for most technologies, even when effects are qualitative aligned with the ones derived with more profiles. As for cost differences, the optimal capacity interval is not a decreasing function of the number of profiles included and stabilizes after enough profiles have been included. More specifically, the interval is strikingly large with 5 profiles and remains significantly so with less than 20 profiles included. One noticeable exception is, again, the stationary Li-ion battery power capacity with smart charging. 

Provided enough profiles are included, qualitative results show that the electrification of the passenger road transport sector has differentiated impacts on optimal capacity and dispatch mixes, which are in line with results previously found in the literature. With smart charging, the introduction of BEVs leads to an increase in wind onshore capacity as well as a slight capacity increase in natural gas technologies. Besides, the impact on optimal solar PV capacity is undetermined and strongly depends on profile selection, no matter the number of BEV profiles included. With bidirectional charging, peak technologies such as natural gas and stationary Li-ion battery see their capacity decrease. At the same time, the capacity of solar PV and wind onshore increases, leading to a greater integration of vRES. 

\vspace{0.25cm}
\subsection{Isolated long trip events shape optimal solutions when too few profiles are included}
\vspace{0.25cm}

\begin{figure}[!ht]
    \centering
    \includegraphics[width=.98\linewidth]{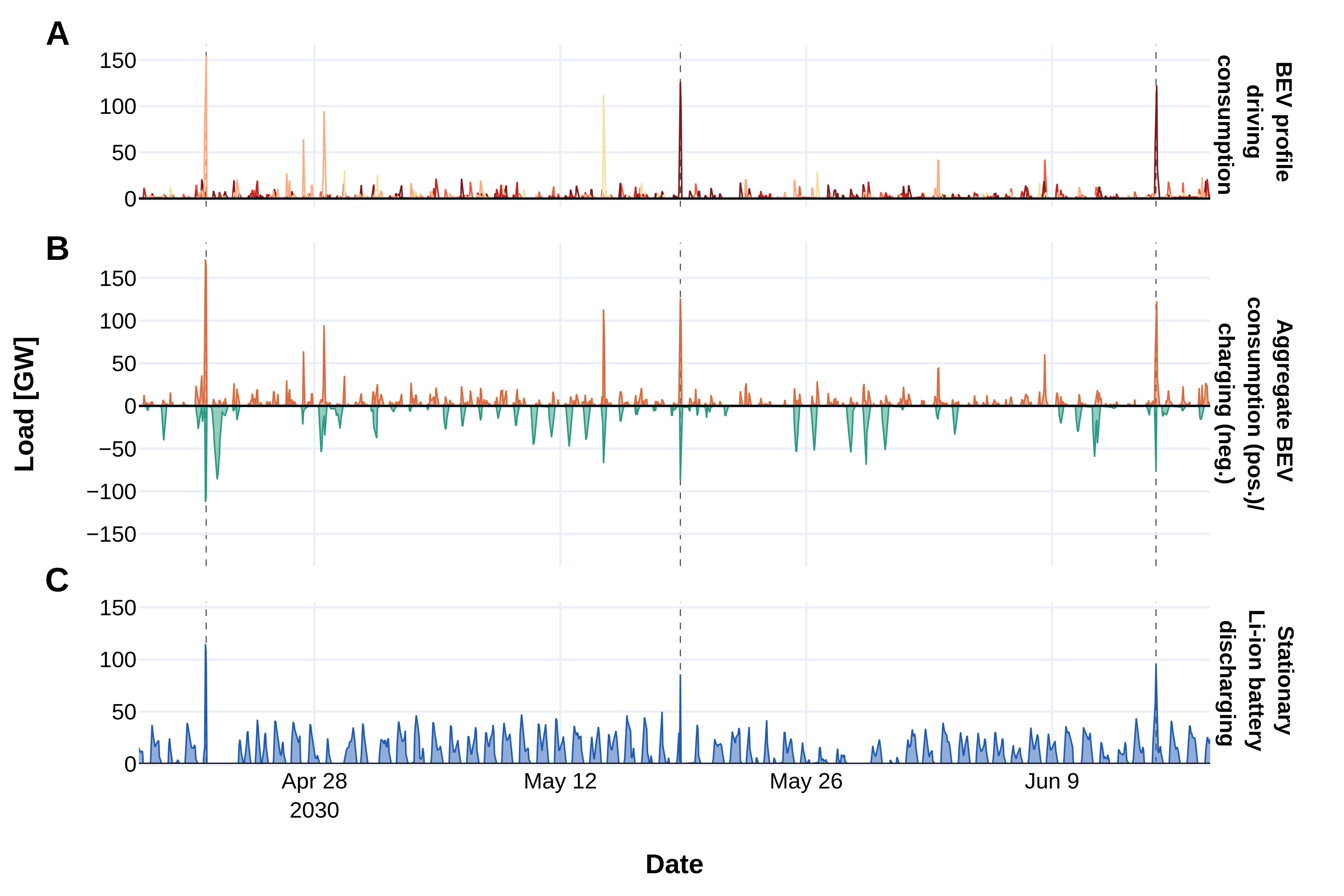}
    \caption{
    \textbf{Selected time series for a scenario with 5 smartly charging BEV profiles and a subset of spring days [Main setting, smart charging].}
    Selected time series are: Individual BEV driving consumption for all 5 profiles (\textbf{A});  
    Aggregate BEV driving consumption (positive values on the y-axis) and aggregate BEV charging (negative values) (\textbf{B}); 
    Stationary Li-ion battery discharging (\textbf{C}).
    }
    \label{fig:timeseries_5profiles}
\end{figure}

When modeling a BEV fleet of constant size with various amounts of BEV profiles, including less BEV profiles entails scaling each profile more. In other words, each BEV profile represents more vehicles or, put differently, more vehicles are assumed to undertake the same trips, charge and discharge, at exactly the same time. In some occurrences, BEV profile generation leads to rather long trips that rely on a larger driving consumption. Scaling up such long mobility events by a large factor results in significantly large spikes for individual BEV driving consumption profiles. When including only 5 BEV profiles to model a 15 million BEV fleet, scaling can result in BEV driving consumption spikes of around 100 to 150 GW at specific hours (Figure~\ref{fig:timeseries_5profiles}, panel A). A driving consumption spike for one individual BEV profile shapes in turn the aggregate BEV driving consumption (Figure~\ref{fig:timeseries_5profiles}, panel B, positive values on the y-axis). 

On the other hand, in order to make these long trips feasible, vehicles must charge during the trip. This hold for all BEV profiles, no matter how much they are scaled, since this derives from the relationship between the total energy needed to undertake the full trip and the BEV battery size. Put differently, BEV batteries are too small to accommodate for such long trips. Nevertheless, when BEV profiles are scaled by a large factor, the larger BEV driving consumption spikes induce larger spikes in BEV charging (Figure~\ref{fig:timeseries_5profiles}, panel B, negative values).     

Such large BEV charging spikes coincide with accordingly large spikes in stationary Li-ion battery discharging (Figure~\ref{fig:timeseries_5profiles}, panel C and Figure~\ref{fig:timeseries_year}, panel A). As more BEV profiles are included, BEV driving consumption spikes are smoothed, as each profile's long trip is scaled with a smaller factor, leading to smoother aggregate BEV driving consumption and BEV charging time series. More particularly, the more profiles are included, the smoother the aggregate BEV driving consumption gets (Figure~\ref{fig:timeseries_evload}). As a consequence, BEV charging as well as stationary Li-ion battery discharging episodes appear to be much less spiky (Figure~\ref{fig:timeseries_year}, panels B and C). 

\begin{figure}[!ht]
    \centering
    \includegraphics[width=.9\linewidth]{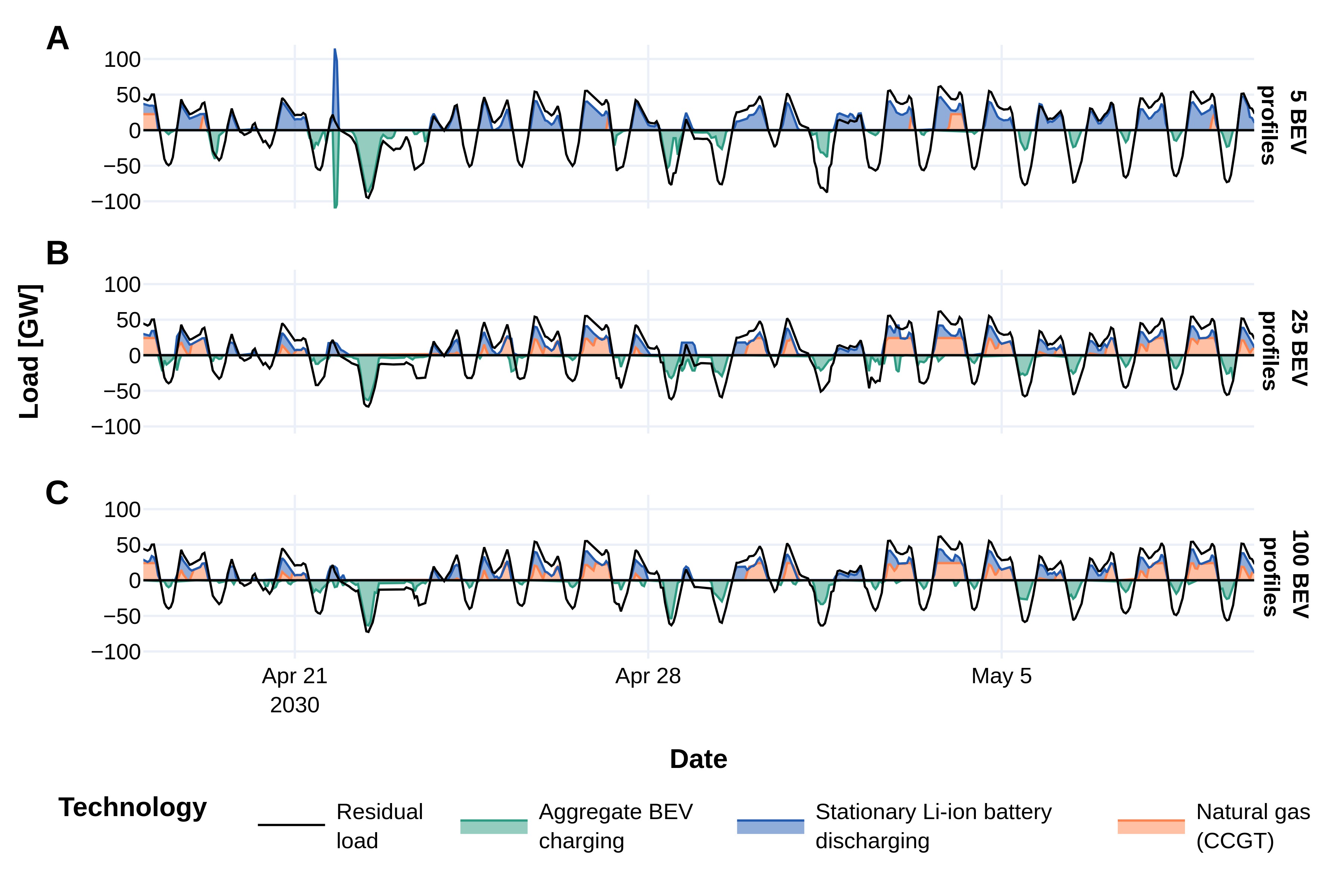}
    \caption{
    \textbf{Optimal dispatch of selected technologies and residual load for a selection of a given number of BEV profiles and a subset of spring days [Main setting].}
    The charging strategy considered is smart charging. The number of profiles displayed is 5 (\textbf{A}), 25 (\textbf{B}) and 100 (\textbf{C}). The residual load computation considers vRES generation after curtailment and does not include the load for BEV charging.
    }
    \label{fig:timeseries_dispatch_smart}
\end{figure}

Long trip episodes are relatively rare over the full year. In the example showcased in Figure~\ref{fig:timeseries_year}, there are around 10 spikes of more than 100 GW aggregate BEV charging when including 5 profiles (panel A), that is to say two long trips per profile per year on average. Rather than their frequency, the magnitude of the spikes themselves is enough to trigger large additional stationary Li-ion battery power capacity investments. To supply these short duration high power spikes, the least-cost solution consists in more battery capacity along with more solar PV, whose diurnal pattern fit short-duration storage more than wind's seasonal ones. On the other hand, the stationary Li-ion capacity increase drives natural gas capacity down, as batteries are needed at full power rate in only a few occasions. In other hours when natural gas technologies would have been dispatched, stationary Li-ion battery is operated, in order to make the most of the large existing capacity (Figure~\ref{fig:timeseries_dispatch_smart}). Hence, the larger battery capacity needed to supply BEV charging spikes driven by isolated long trip events displaces gas capacity, in line with what has been described previously. When more profiles are included, long trip events do not lead to BEV charging spikes that are large enough to trigger additional Li-ion battery power capacity. In these cases, the battery capacity in the existing mix does not suffice to supply the demand in periods of high residual load, which requires natural gas technologies. 

For bidirectional charging, the same logic as with smart charging is at play. However, since part of the BEV fleet can discharge to supply BEV charging spikes (Figure~\ref{fig:timeseries_dispatch_v2g}), the additional Li-ion battery capacity investment needs are smaller and, as the number of profiles grows and the magnitude of spikes decreases, the BEV discharging capacity is large enough to supply them entirely. More generally, the BEV discharging capacity is large enough to also displace part of capacity of natural gas technologies while integrating more RES.

\subsection{A sufficiently high number of profiles is also important when modeling smaller fleets}
\vspace{0.25cm}

Scaling BEV profiles plays a paramount role in shaping BEV consumption spikes, which decisively drive optimal capacity and dispatch results. However, the scaling factor depends on both the number of BEV profiles included and the total size of the BEV fleet that is represented. It is not clear how these two variables play against each other. When modeling a smaller BEV fleet, less profiles could be needed as each profile would represent less vehicles, and conversely, a larger BEV fleet may require more profiles. We analyze how robust our findings are for BEV fleets of 5, 10 and 20 million vehicles. In order to ease comparability across fleet sizes, we use the number of BEVs per BEV profile as a common metric since it captures the interplay between, on the one hand, the number of profile and, on the other, the total BEV fleet size. For a given fleet size, the correspondence between the number of included BEV profiles and the number of BEVs per BEV profile is detailed in Table~\ref{tab:number_bevs_profiles}.

\begin{table}[!ht]
\centering
\begin{tabular}{@{}cccccccc@{}}
\toprule
\multicolumn{1}{l}{} & \multicolumn{1}{l}{} & \multicolumn{6}{c}{\textbf{\begin{tabular}[c]{@{}c@{}}Number of BEVs per BEV   profile \\ {[}in thousands{]}\end{tabular}}} \\ \midrule
\multicolumn{1}{l}{} &
  \multicolumn{1}{l}{\textbf{}} &
  \textbf{1,000} &
  \textbf{500} &
  \textbf{250} &
  \textbf{200} &
  \textbf{143} &
  \textbf{125} \\ \midrule
\multirow{4}{*}{\textbf{\begin{tabular}[c]{@{}c@{}}BEV fleet size \\ {[}in million{]}\end{tabular}}} &
  \textbf{5} &
  5 &
  10 &
  20 &
  25 &
  35 &
  40 \\
 &
  \textbf{10} &
  10 &
  20 &
  40 &
  50 &
  70 &
  80 \\
 &
  \textbf{15} &
  15 &
  30 &
  60 &
  75 &
  105 &
  120 \\
 &
  \textbf{20} &
  20 &
  40 &
  80 &
  100 &
  140 &
  160 \\ \bottomrule
\end{tabular}
\caption{
\textbf{Number of included BEV profiles for a given BEV fleet size and a given number of BEVs per BEV profile.}
}
\label{tab:number_bevs_profiles}
\end{table} 

\begin{figure}[!ht]
    \centering
    \includegraphics[width=0.9\linewidth]{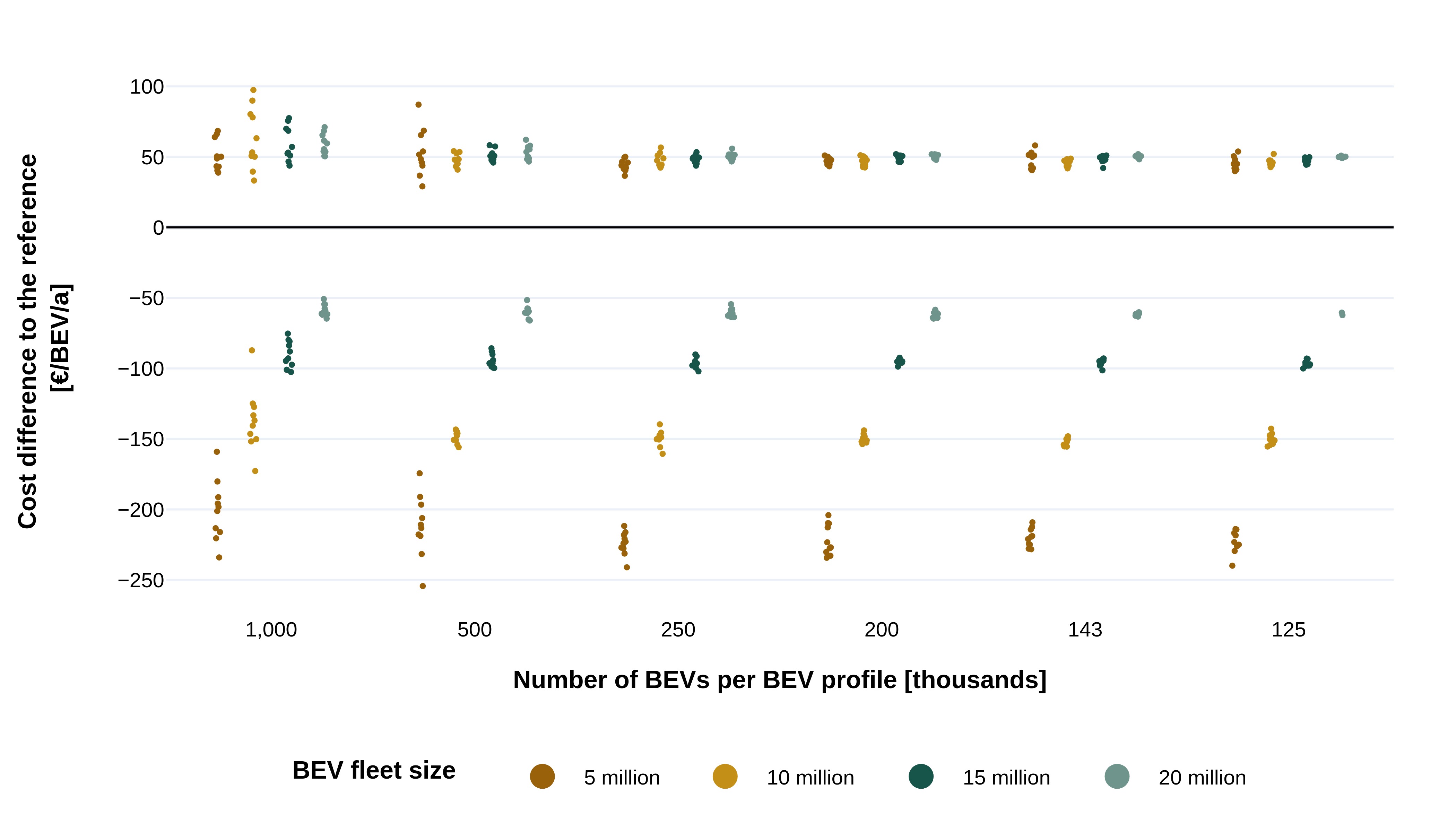}
    \caption{
    \textbf{System cost difference to the reference by number of BEVs per profile, fleet size and charging strategy.} Scenarios with smart (resp. bidirectional) charging exclusively have positive (resp. negative) values. Differences are expressed relative to the reference with no BEVs. For $x$ a given number of BEV profiles, each dot refers to the optimal solution of a model including a randomly drawn sample of $x$ BEV profiles. \textit{Note}: for a 20 million BEV fleet, 125,000 BEVs per profile and bidirectional charging, only 2 scenarios are displayed, as the other 8 samples did not solve within the specified time limit.
    }
    \label{fig:systemcosts_fleetsizes}
\end{figure}

For a smaller fleet of 5 million BEVs, representing more than 250,000 BEVs per profile leads to a noticeable variation across samples in cost differences to the reference for both charging strategies (Figure~\ref{fig:systemcosts_fleetsizes}). In other words, including only 5 or 10 BEV profiles make cost results depend quite strongly on profile selection, just as in the case of a bigger fleet of 15 million BEVs. When including 20 profiles, which corresponds to 250,000 BEVs per profile, results do not vary much across samples and are in line with results obtained when including more profiles, i.e., when decreasing the number of BEVs represented by each BEV profile. 

A similar conclusion can be drawn for a fleet size of 10 million BEVs, as 500,000 BEVs per profile, which corresponds to 20 BEV profiles in this case, ensures cost results stability for both strategies. On the other hand, for a larger fleet of 20 million BEVs, cost results prove to be already quite stable with each BEV profile representing one million cars, that is to say when including 20 BEV profiles in total.

Nevertheless, as in the case with 15 million BEVs, cost results show a smaller sensitivity to the overall number of profiles than capacity results (Figure~\ref{fig:capacity_fleetsizes}). This is especially the case with smart charging for stationary Li-ion battery, where optimal capacity can vary significantly even when cost results are rather robust, e.g., with one million BEVs per profile for a 20 million BEV fleet, or with 500,000 BEVs per profile in a 10 million fleet. Results prove to be more robust when including enough profiles such that each profile represents at most 250,000 BEVs. Robustness is further improved for 200,000 BEVs per profile but this might incur unrealistic runtimes when modeling larger BEV fleets, as this corresponds to, e.g., 100 profiles for a 20 million BEV fleet. 

\section{Discussion}
\label{sec:discussion}

Based on our findings, holding together both runtime and accuracy considerations, we derive a rule of thumb, which consists in including enough BEV profiles such that each profile represents between 250,000 and 200,000 BEVs. For a 15 million BEV fleet, this corresponds to including between 60 and 75 BEV profiles. While including less profiles (about 20) already ensures robust cost results for most fleet sizes and settings, increasing the number of profiles to, e.g., 60 or~75 in the 15~million fleet case, does not come at a great additional runtime expense. Even in a more computationally demanding setting with interconnected neighboring countries, a problem including 60~profiles can be solved in around 45~minutes with smart charging and a bit less than 90~minutes with bidirectional charging. On the other hand, these additional profiles do provide a greater accuracy when it comes to capacity results, in particular when analyzing smart charging and Li-ion battery capacity. 

Previously mentioned practical guidelines and findings should not be applied nor interpreted without some key limitations in mind. First, it is important not to over-interpret how optimal solutions' intervals compare for various amounts of BEV profiles. Indeed, all samples are drawn from a common pool of BEV profiles whose size is constant and amounts to 200. Hence, for larger amounts of profiles, samples are more likely to be more similar one to another. In other words, the probability that two samples including the same number of profiles have one or several profiles in common increases. This naturally leads to a decrease in the variability in optimal solutions across samples as the number of profiles increases. That being said, our results show a stagnating interval rather than a monotonically decreasing one. While we are confident that our result interpretation holds, a sensitivity analysis consisting in enlarging significantly the initial pool of BEV profiles would help strengthen our conclusions. 

Furthermore, absolute runtime values depend on the setting considered and, as importantly, on the performance characteristics of the hardware used. Hence, they should not be taken at face value and should be interpreted cautiously, as runtimes could be decreased by using better performing hardware. Yet, as an hardware downgrade or upgrade would probably apply uniformly to all problems, it would not qualitatively change our findings but could lead to consider problems with more or less profiles. Related to this question, our current results would benefit from further investigating why runtimes are somewhat non-monotonic in the total number of profiles, in particular when more than 100~BEV profiles are included.

We purposefully design an initial pool of BEV profiles with various BEV characteristics. On the one hand, this makes our results more applicable to settings aiming at a realistic representation of the vehicle fleet, and hence more easily generalizable. On the other, this leads to various aggregate BEV characteristics for a given sample and limits, to some extent, the comparability of results across samples. While we conduct robustness checks by excluding samples deviating from average aggregate characteristics by more than~5\%, it would be of interest to isolate the pure effect of the amount of BEV profiles by only drawing profiles that share the same characteristics. Drawing more than 10 samples for a given number of profiles could also ease comparing optimal solution intervals across various amounts of BEV profiles.

Apart from implementing previously mentioned sensitivity checks, comparing how results from different modeling approaches differ would add value to this research stream. While this study specifically looks at one approach where (potentially many) individual BEV profiles are included, other studies assess the flexibility potential of battery electric vehicles by including only one profile, resulting from the previous aggregation of many mobility profiles. It would also be interesting to understand whether average optimal solutions from a large number of samples including less profiles would converge to the optimal solution of a problem with more profiles. If yes, understanding how many samples should be drawn and analyzing whether this method enables runtime gains would be of practical interest. In particular, this would give useful insights for computationally demanding applications with many nodes and sector coupling options that may accommodate for only a limited number of profiles. It is indeed not clear at this stage whether including, e.g., 60~BEV profiles in all neighboring power sectors would still be tractable. Benchmarking the variability range in optimal results for different BEV profile samples against the variability induced by weather year selection would also help put absolute values in perspective and decide which sensitivity checks must be prioritized if needed. Lastly, more interdisciplinary work between energy system modelers and transport modelers needs to be carried out to understand how to retain a representative profile selection when only a few dozens can be included in power sector models.   

\section{Conclusion}

Including individual BEV profiles in power sector models increases the complexity of cost-minimization problems and introduces a new layer of sensitivity in the results that needs to be better understood. 

First, our findings highlight that including too few BEV profiles inflates the estimated costs of coupling the passenger road transport and the power sectors. Having too few profiles also leads to distorted capacity and dispatch results, particularly for stationary Li-ion battery and natural gas turbine technologies. This is driven by isolated long-trip events that trigger very large BEV charging spikes.

Second, we find that cost results are less sensitive than capacity and dispatch results. Similarly, capacity and dispatch outcomes for smart charging are more sensitive than those for bidirectional charging. 

Third, our study shows that the impact of profile selection, i.e., drawing alternative BEV profiles for a given number of profiles, is negligible when enough BEV profiles are included. One noticeable exception is the optimal capacity for solar PV with smart charging. In this case, absolute capacity results should be interpreted cautiously. 

Fourth, we emphasize that including a large number of BEV profiles may be inefficient since, beyond a certain point, adding more profiles increases runtime without significantly reducing the variation in optimal solutions across samples. Considering runtime and accuracy, we recommend including enough profiles such that each BEV profile represents 200,000 to 250,000 BEVs when modeling BEV fleets of 5 to 20 million vehicles. When computational constraints prevent the inclusion of sufficiently many profiles, we strongly recommend to run sensitivity checks on profile selection for two specific technologies: stationary Li-ion batteries and natural gas turbines.

We conclude that future research related to the impact of battery electric vehicles on the power sector should focus on two specific points. First, understand how the two main modeling approaches, the aggregate approach and the individual profile approach, compare and influence optimal results. Second, gain insights into the best way to design a relatively small pool of BEV profiles that remains representative of nationwide electric car mobility behaviors. 

\section{Methods}
\label{sec:methods}

\subsection{Battery electric vehicle profiles}

Each battery electric profile corresponds to a set of three time series generated with the open-source stochastic tool \textit{emobpy}\autocite{gaete2021open}. Using empirical distributions from the national representative travel survey \textit{Mobilität in Deutschland}\autocite{eggs2018mobilitat}, \textit{emobpy} first enables to generate a \textit{mobility} time series at a 15-minute resolution for the full representative year. It corresponds to a time series of car trips, i.e. when, how long and to which destination a car is driven. A second time series, the \textit{driving electricity consumption} time series, can be derived from the mobility time series assuming BEV characteristics such as size and weight of the vehicle. Finally, it allows to generate a third \textit{grid availability} time series, which specifies when and at which power rating a vehicle is connected to a charging station. This last time series relies on assumptions on the distribution of charging stations at each destination. 

Possible trip destinations are workplace, shopping, errands, escort, leisure, and home. Four BEV models are considered (Model 3 (Tesla), ID.3 (VW), Kona (Hyundai) and Zoe (Renault)) and parametrized using 2021 data from the Electric Vehicle Database\autocite{evdatabase}. For a more detailed explanation of how these time series relate to each other and what assumptions are made, please refer to Gaete et al. (2021)\autocite{gaete2021open}. \\

Based on a common pool of 200 profiles generated with \textit{emobpy}, we randomly draw profiles to build samples. A sample's size ranges from 5 to 120 BEV profiles at an increment of 5. For a given sample size, we generate ten random samples. Within a sample, each profile can only appear once. After each sample is complete, profiles are replaced in the common pool, meaning that different samples can feature identical BEV profiles.

\begin{figure}[!ht]
    \centering
    \includegraphics[width=1\linewidth]{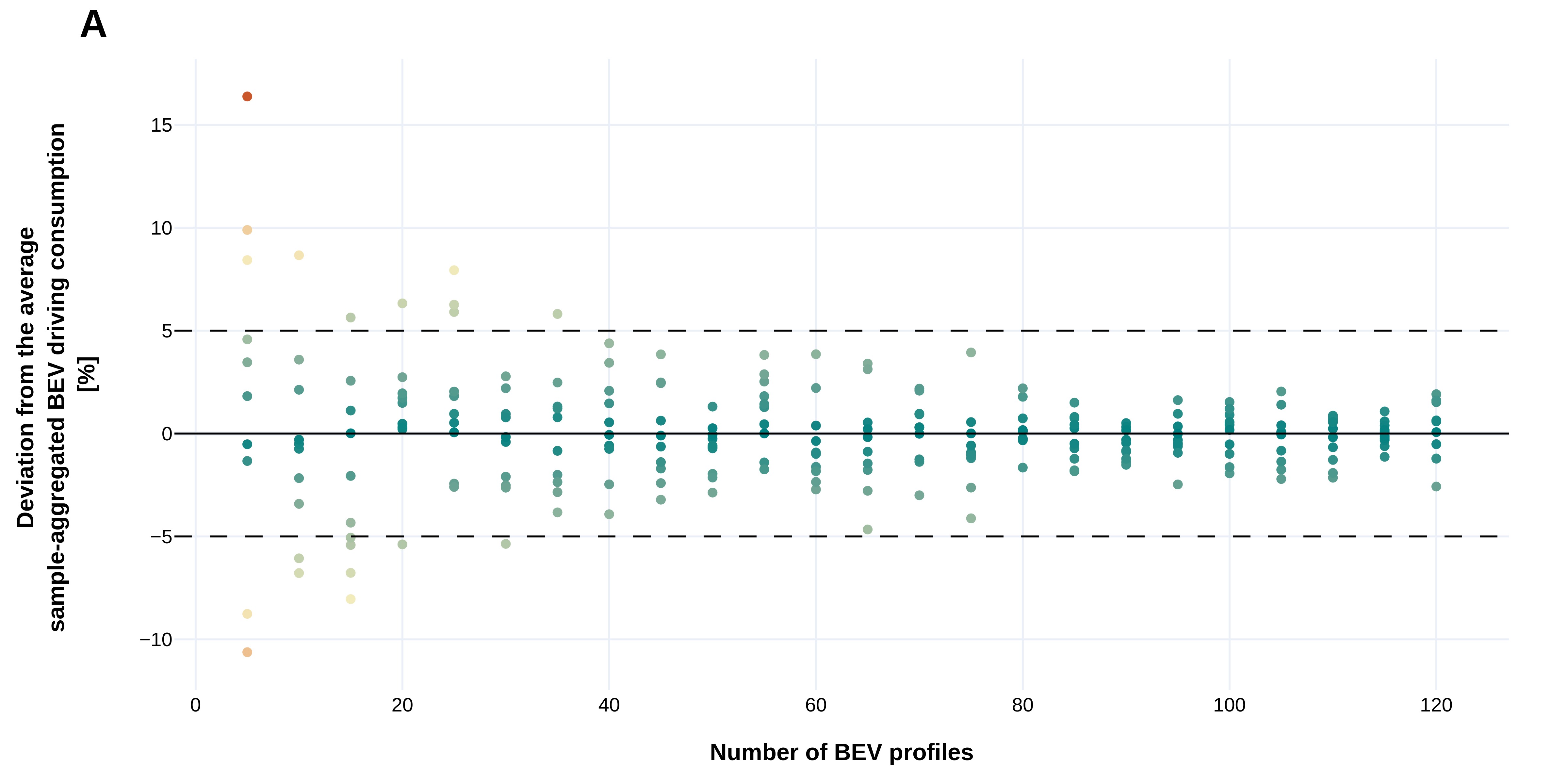}
    \includegraphics[width=1\linewidth]{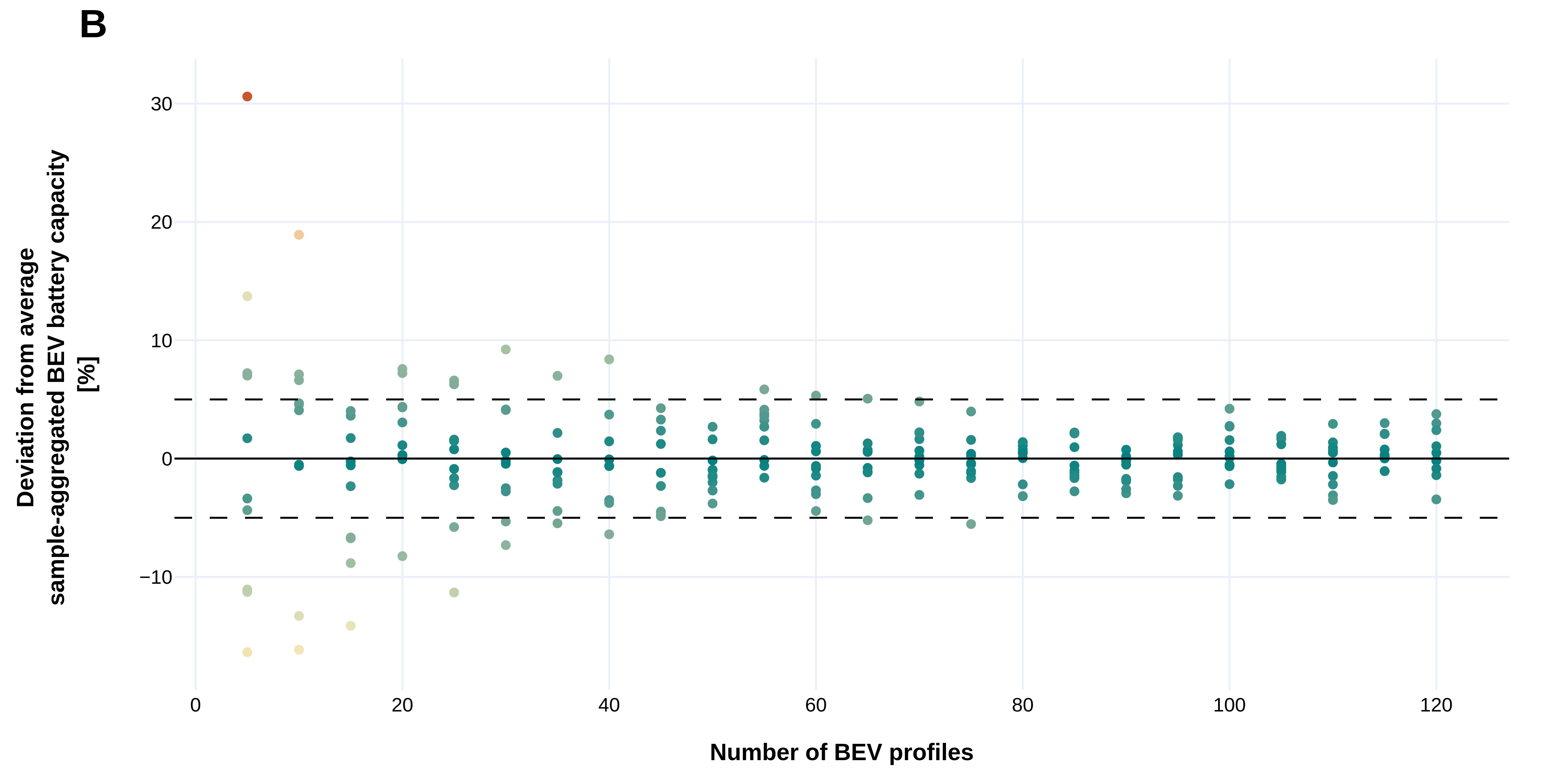}
    \caption{
    \textbf{Sample deviation to the average aggregate BEV characteristics by number of BEV profiles.} 
    Aggregate BEV characteristics refer to the BEV electricity driving consumption aggregated over all scaled BEV profiles within each sample (\textbf{A}) and, similarly, to the BEV battery capacity aggregated over all scaled BEV profiles within each sample (\textbf{B}). For each aggregate characteristic, the average is then computed across all samples. Dashed lines delimit the 5\% corridor around the average. The color scale reflects the distance to the average value and not the sample ID. Samples with similar colors in panels A and B are not necessarily identical.
    }
    \label{fig:bev_delta_aggregates}
\end{figure}

Randomly drawing profiles to create samples entail that samples differ one from another when it comes to aggregate BEV characteristics such as the BEV battery capacity or the total driving electricity consumption, even across samples of equal size, since profiles reflect different BEV models and trip sequences. For a 15 million BEV fleet, the aggregate BEV battery capacity for a sample varies from 734 GWh to 1,146 GWh (Figure~\ref{fig:bev_characteristics_absolute_consumption_capacity}, panel A). This corresponds to a maximum sample average of 76 kWh per BEV (respectively, a minimum sample average of 49 kWh), with an average aggregate of 58 kWh per vehicle across all samples. Similarly, the yearly aggregate driving electricity consumption varies across samples between 29 TWh and 38 TWh (Figure~\ref{fig:bev_characteristics_absolute_consumption_capacity}, panel B), i.e. a sample average varying from 1.9 MWh and 2.5 MWh per vehicle per year, with an average of 2.2 MWh per vehicle per year across all samples. Even if BEV aggregate characteristics vary across samples, most samples show comparable aggregates for the two aforementioned parameters, as most of them fall within a 5\% corridor around the average value (Figure~\ref{fig:bev_delta_aggregates}). 

\vspace{0.25cm}
\subsection{Power sector model}
\vspace{0.25cm}

We integrate BEV profile samples into the open-source power sector model DIETER\footnote{This work relies on the julia implementation of DIETER, accessible on a GitLab public repository at the following link: \href{https://gitlab.com/diw-evu/dieter_public/DIETERjl}{https://gitlab.com/diw-evu/dieter\_public/DIETERjl}.} (\textit{Dispatch and Investment Electricity Tool with Endogenous Renewables})\autocite{gaete2021dieterpy, zerrahn2017long}. It is a linear model minimizing with perfect foresight investment and dispatch costs over the 8,760 consecutive hours of a representative year for various electricity generation and storage technologies, under technical and feasibility constraints. This model has been previously used to quantify power sector impacts of sector coupling technologies - such as heat pumps\autocite{schmidt2025mix, roth2024power}, electrolyzers for green hydrogen production\autocite{kirchem2023power, stockl2021optimal} or battery electric vehicles\autocite{gueret2025moderate, gueret2024impacts} - and flexibility requirements in future power systems based on variable renewables\autocite{kittel2024coping, roth2023geographical}. 

\subsubsection*{Main and alternative settings}

Results put at the forefront of this paper rely on a Germany as an island framework with a 2030 horizon. This main framework does not include sector coupling technologies except for battery electric vehicles in scenarios. When modeling the integration of BEVs, we assume a 15 million vehicle fleet. 

We also put results for the main setting in perspective and design two alternative settings. The first alternative setting includes, on top of BEVs, an additional industrial demand for green hydrogen, amounting to 30 TWh a year and assumed to be equally distributed over all hours of the year. The second alternative setting includes interconnections with Germany's neighboring countries (Austria, Belgium, Czech Republic, Denmark, France, Luxembourg, the Netherlands, Poland, Switzerland) and Italy. The power sector of each considered country is explicitly modeled as well as net transfer capacities (NTC) across countries (Table~\ref{tab:parameters_ntc}). 

Hence, there is one main reference referring to the main setting and two alternative references referring each to the previously introduced alternative settings. For each setting, scenarios differ from their respective references only by the additional inclusion of electric vehicles. 

\subsubsection*{Technologies and capacity constraints}

We consider both renewable and conventional generation technologies. Renewable technologies include bioenergy, run-of-river, solar photovoltaic, wind onshore and wind offshore. Conventional technologies include lignite, hard coal, natural gas (OCGT and CCGT), nuclear, oil and a residual other fossil fuel technology. Additionally, we include stationary Li-ion batteries, pumped-hydro (closed and open), hydro reservoirs as storage technologies. A combination of compressed green hydrogen stored in caverns and reconverted to electricity via H$_2$-ready gas turbines is introduced as a long-duration storage technology. 

For Germany, capacity expansion for solar photovoltaic, wind onshore and natural gas technologies (OCGT and CCGT) is left unconstrained. Wind offshore is constrained to a maximum of 30 GW out of policy relevance considerations and nuclear is fixed to zero. Hydropower and bioenergy capacity are fixed, based on data from the ENTSO-E Pan-European Climate Database (PECD 2021.3)\autocite{de2022entso}, since these technologies notably have a very limited expansion potential. 

In the alternative setting including neighboring countries and Italy, the parametrization does not change for Germany. For all other countries, capacity in renewable technologies and nuclear are fixed, using data from a Ten-Year Network Development Plan (TYNDP) of ENTSO-E\autocite{entsoe_tyndp_2018}. The investment in other dispatchable technologies is left endogenous within capacity bounds. Table \ref{tab:parameters_capacitybounds} summarizes country-specific capacity constraints for all considered technologies. The dispatch for all technologies is endogenous and unconstrained. 

In order to specifically assess the robustness of runtime results (Figure~\ref{fig:runtimes_stylized_othersettings}, panel A), we also design a very stylized setting for Germany as an island with only five generation technologies (CCGT, OCGT, offshore wind, onshore wind and solar photovoltaic), for which all capacity bounds are removed, and no storage technologies. 

\subsubsection*{Cost, weather and demand data}

Table~\ref{tab:parameters_generation} and Table~\ref{tab:parameters_storage} detail cost parameters for generation and storage technologies. These parameters are mostly taken from the Danish Energy Agency\autocite{dea_techdata2024}. We assume that a carbon price of 130~euros per tCO$_2$eq adds up to the operational costs of fossil generation technologies. Transmission and distribution network costs are not considered.

We derive capacity factors time series for variable renewable technologies (solar, wind onshore, wind offshore) from the ENTSO-E Pan-European Climate Database (PECD 2021.3)\autocite{de2022entso}, using 2008 as a weather year.  Inflow time series for run-of-river, reservoirs and open pumped-hydro come from the same source. 

Electricity demand data for all countries (Table \ref{tab:parameters_load}) derive from the ENTSO-E Pan-European Climate Database (PECD 2021.3)\autocite{de2022entso}, using estimates for 2030 and the weather year 2008. In Germany, the assumed yearly load amounts to 583~TWh.

\subsection{Software and hardware characteristics}

We use Gurobi as a solver implementing the barrier method with no crossover. Computations are run on a High Performance Cluster (HPC). To ensure comparability of runtime results, we only select nodes with similar central processing unit (CPU) and random access memory (RAM) characteristics. Nodes used have two CPUs of type AMD EPYC 7302, with 16 cores and 32 threads. Each node has a RAM of 256 gigabytes. 

\section*{Data and code availability}

The code for data and results processing as well as all input data are stored in a public GitLab repository available at \href{https://gitlab.com/diw-evu/projects/bevs-in-psm}{this link}. DIETERjl is publicly available on \href{https://gitlab.com/diw-evu/dieter_public/DIETERjl}{GitLab}. The python package for \textit{emobpy} is available at \href{https://pypi.org/project/emobpy/}{https://pypi.org/project/emobpy/} and a \href{https://emobpy.readthedocs.io/en/latest/index.html}{documentation page} provides further instructions. MiD data are also publicly available and can be retrieved at \href{https://www.mobilitaet-in-deutschland.de/archive/MiT2017.html}{https://www.mobilitaet-in-deutschland.de/archive/MiT2017.html}. 

\section*{Acknowledgments}

We thank Carlos Gaete-Morales, Dana Kirchem, Martin Kittel, Alexander Roth, Felix Schmidt and Wolf-Peter Schill for developing and maintaining the open-source DIETER model and \textit{emobpy} stochastic tool. We also thank Wolf-Peter Schill for his useful feedback on some preliminary results. We are grateful to Norbert Paschedag for excellent support related to TU Berlin's High Performance Cluster. We acknowledge a research grant by the German Federal Ministry of Education and Research via the ``Ariadne'' project (Fkz 03SFK5NO-2).  

\section*{Declaration of interests}

The author declares no competing interests.

\printbibliography

\appendix
\setcounter{figure}{0}
\renewcommand{\thefigure}{SI.\arabic{figure}}
\setcounter{table}{0}
\renewcommand{\thetable}{SI.\arabic{table}}
\renewcommand{\thesubsection}{SI.\arabic{subsection}}

\clearpage

\section*{Supplemental Information}
\subsection{Additional graphs}

\begin{figure}[!ht]
    \centering
    \rotatebox{90}{\includegraphics[width=1.05\linewidth]{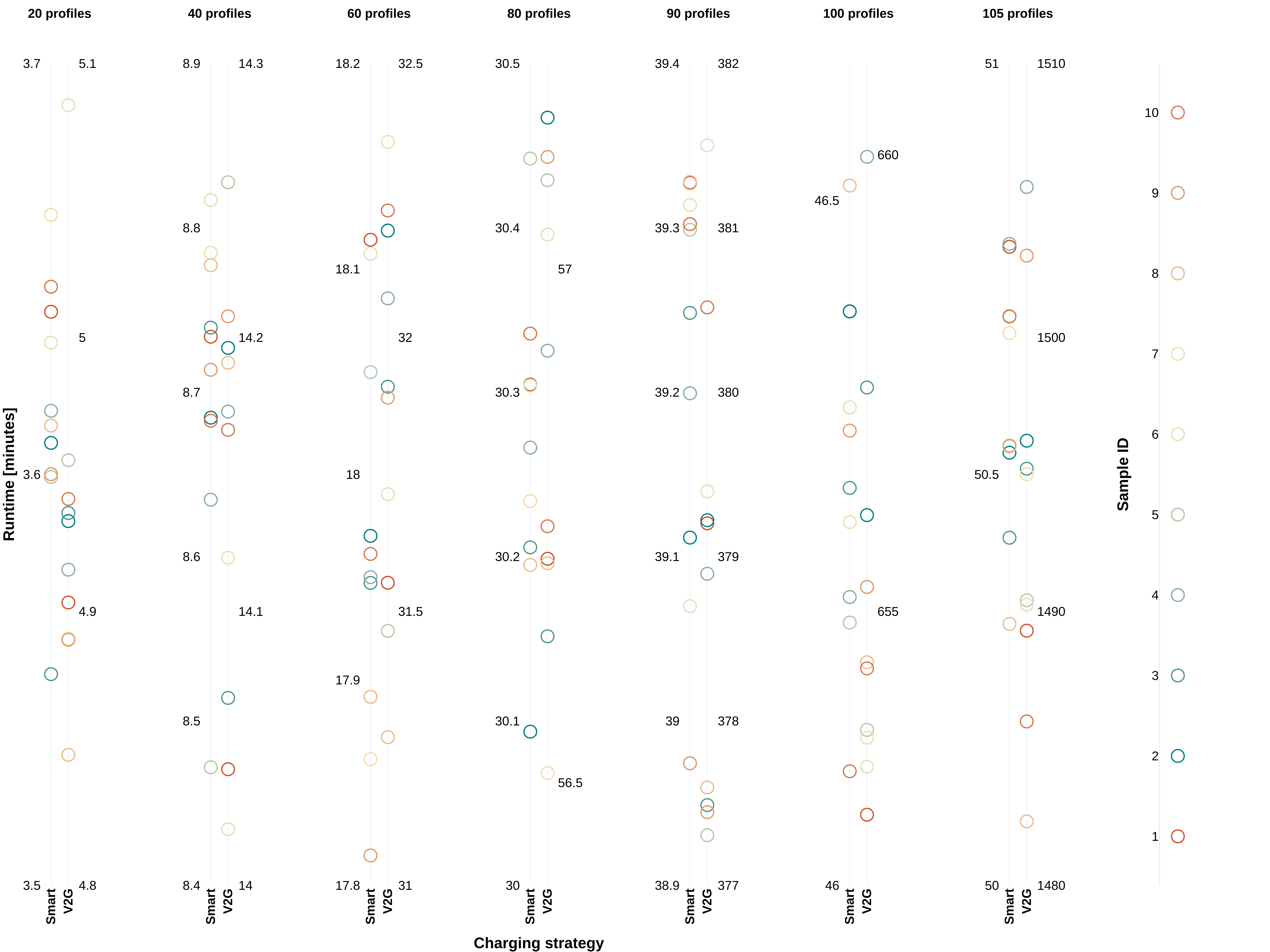}}
    \caption{
    \textbf{Runtimes for a selected number of BEV profiles by sample ID and charging strategy [Main setting]}.
    Runtime encompasses all steps from loading data to model computation and solving. For each number of BEV profiles and charging strategy, the scale is adjusted so as to display as distinctively as possible each sample. 
    }
    \label{fig:runtimes_samples}
\end{figure}

\begin{figure}[!ht]
    \centering
    \includegraphics[width=1\linewidth]{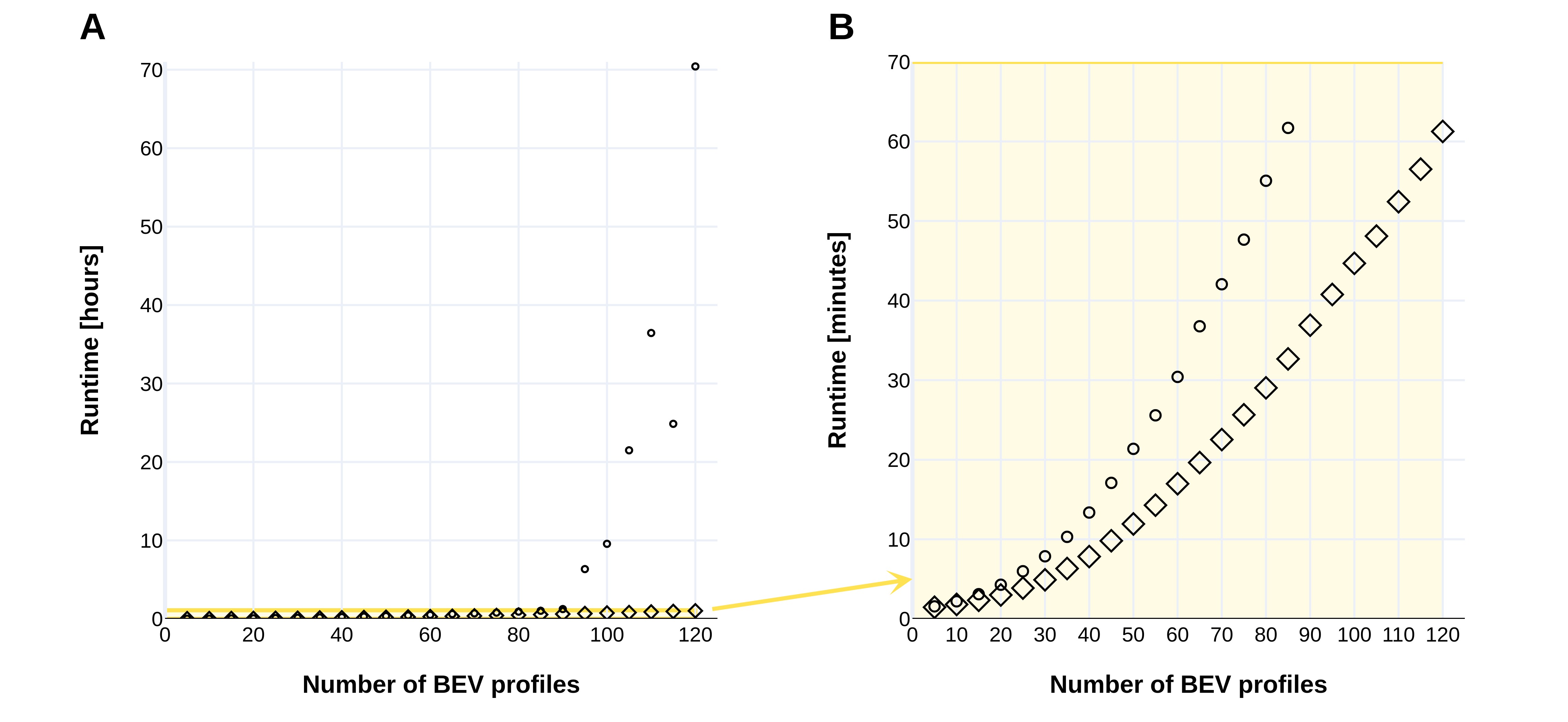}
    \includegraphics[width=1\linewidth]{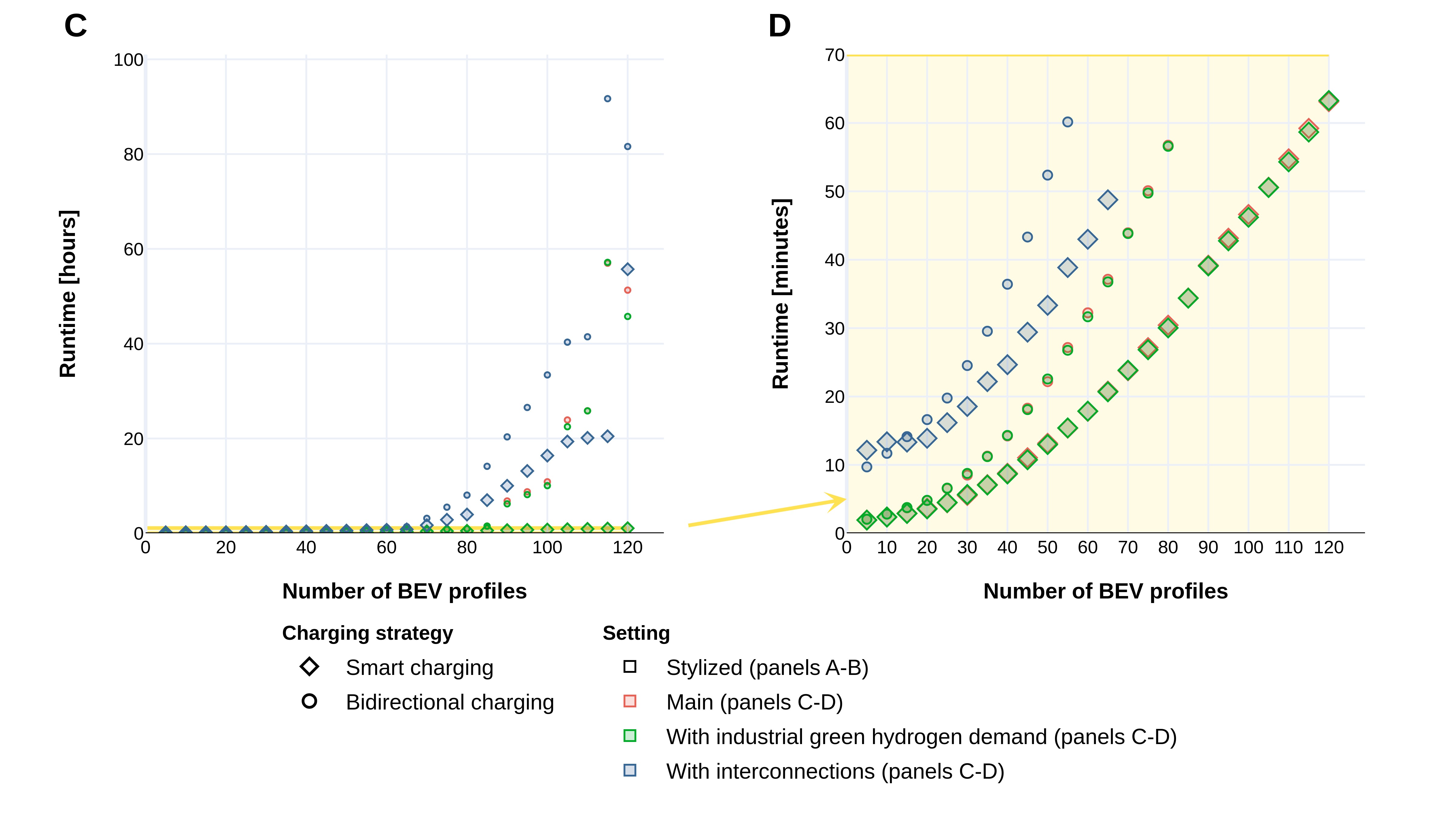}
    \caption{
    \textbf{Runtime by number of BEV profiles and charging strategy for alternative settings.} 
    Runtime encompasses all steps from loading data to model computation and solving. Runtime in minutes for a selection of scenarios (\textbf{A, C}). Zoom on runtimes up to 70 minutes (\textbf{B, D}). Alternative settings include: a stylized setting (\textbf{A-B}), a setting with industrial green hydrogen demand and a setting with interconnected neighboring countries (\textbf{C-D}). Runtime for the main setting are also displayed on panels C and D.
    }
    \label{fig:runtimes_stylized_othersettings}
\end{figure}

\clearpage

\begin{figure}[!ht]
    \centering
    \includegraphics[width=1\linewidth]{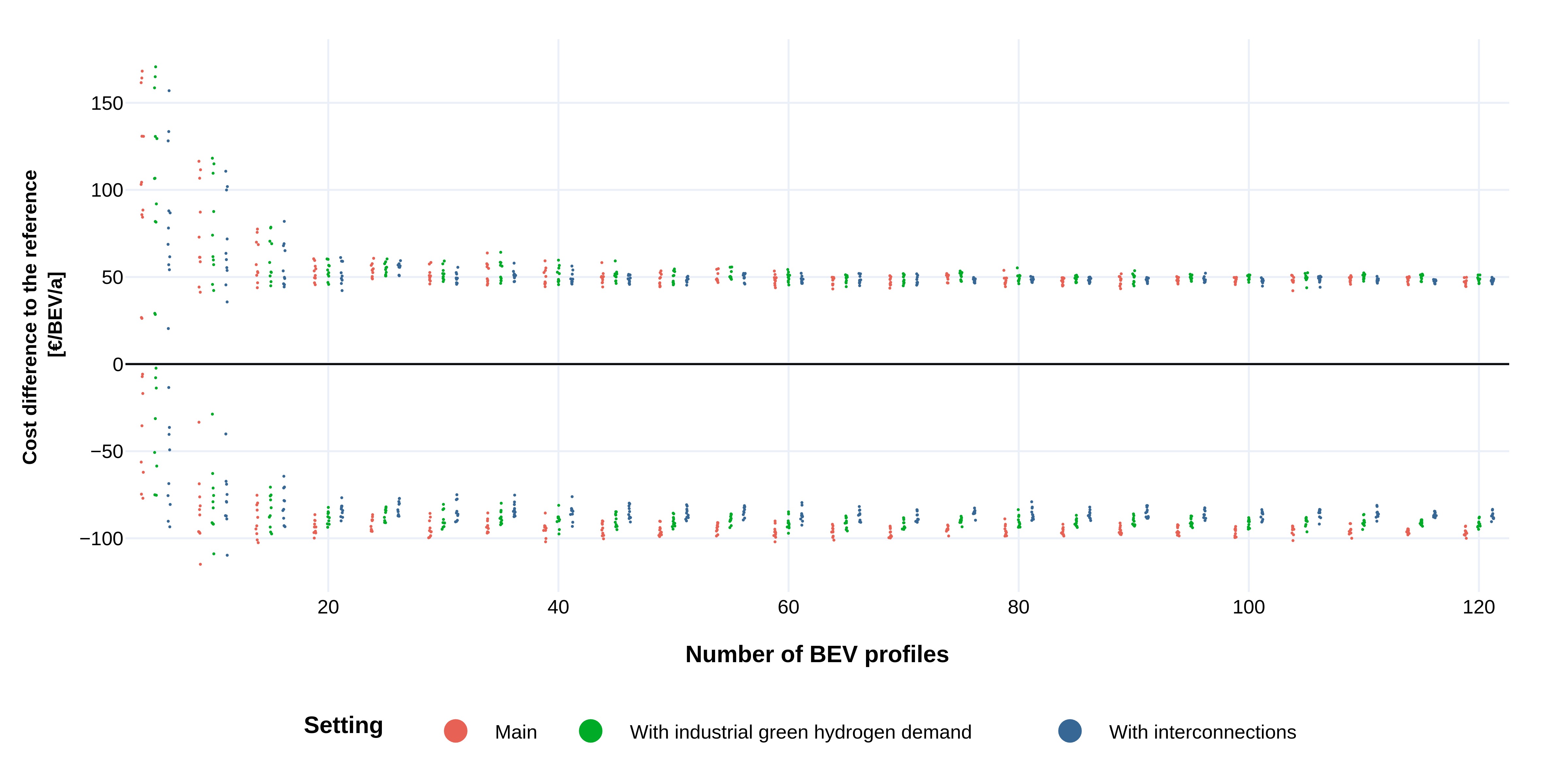}
    \caption{
    \textbf{System cost difference to the reference by number of BEV profiles and charging strategy [Alternative settings].}
    Differences are expressed relative to the reference with no BEVs. For $x$ a given number of BEV profiles, each dot refers to the optimal solution of a model including a randomly drawn sample of $x$ BEV profiles. Optimal outcomes for smart charging have exclusively positive values. Respectively, outcomes for bidirectional charging have almost all negative values, except for the case with 5 profiles, where one sample in each setting has a positive value (the lowest positive value).}
    \label{fig:systemcosts_settings}
\end{figure}

\begin{figure}[!ht]
    \centering
    \includegraphics[width=1\linewidth]{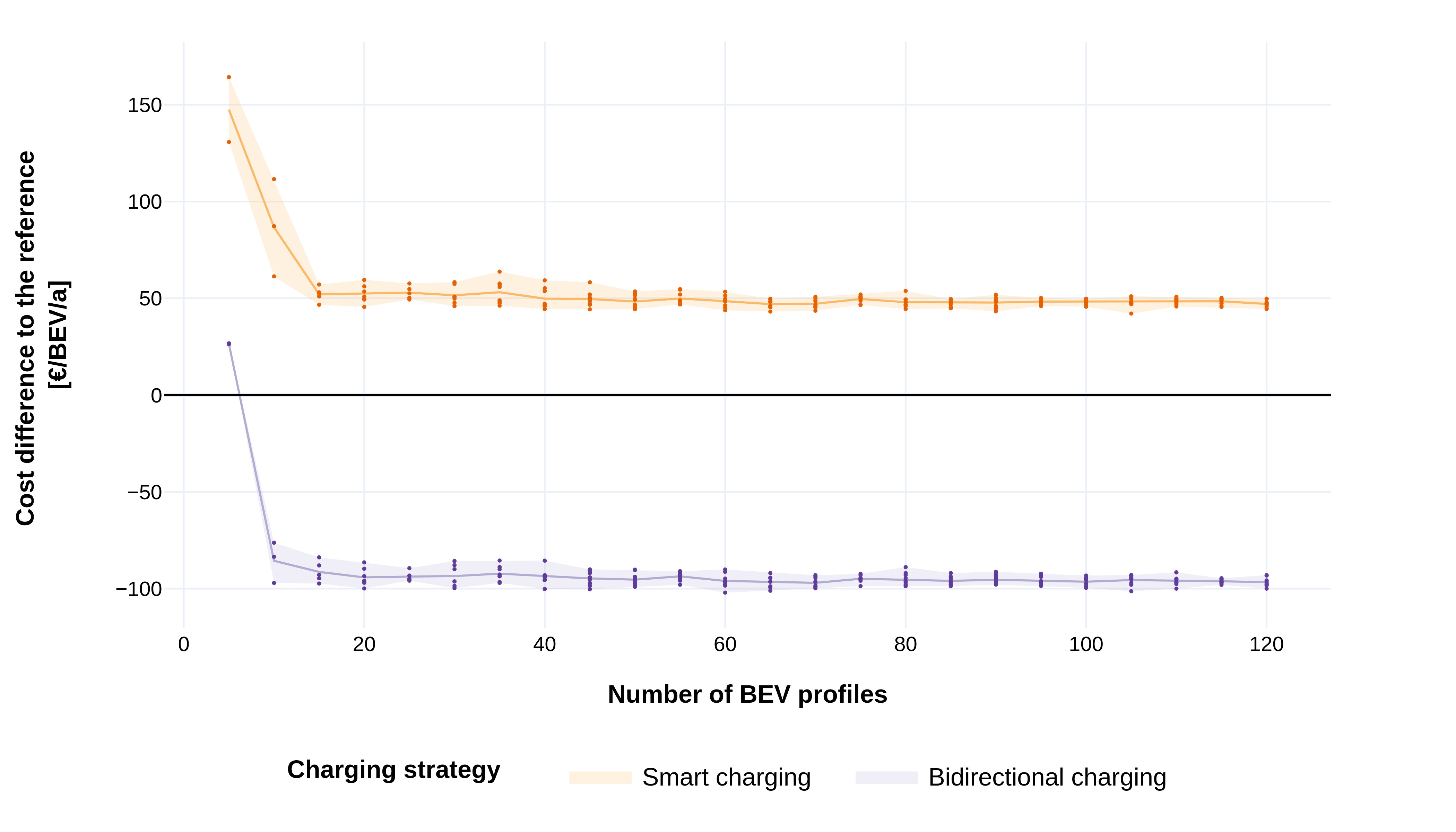}
    \caption{
    \textbf{System cost difference to the reference by number of BEV profiles and charging strategy [Main setting, trimmed].}
    Only scenarios whose aggregate BEV characteristics do not deviate from the average by more than 5\% are displayed. Differences are expressed relative to the reference with no BEVs. For $x$ a given number of BEV profiles, each dot refers to the optimal solution of a model including a randomly drawn sample of $x$ BEV profiles. Lines connect sample averages for a given charging strategy. Shaded areas delimit the empirical cost difference interval for a given charging strategy.}
    \label{fig:systemcosts_trimmed_both}
\end{figure}

\clearpage

\begin{figure}[!ht]
    \centering
    \includegraphics[width=0.95\linewidth]{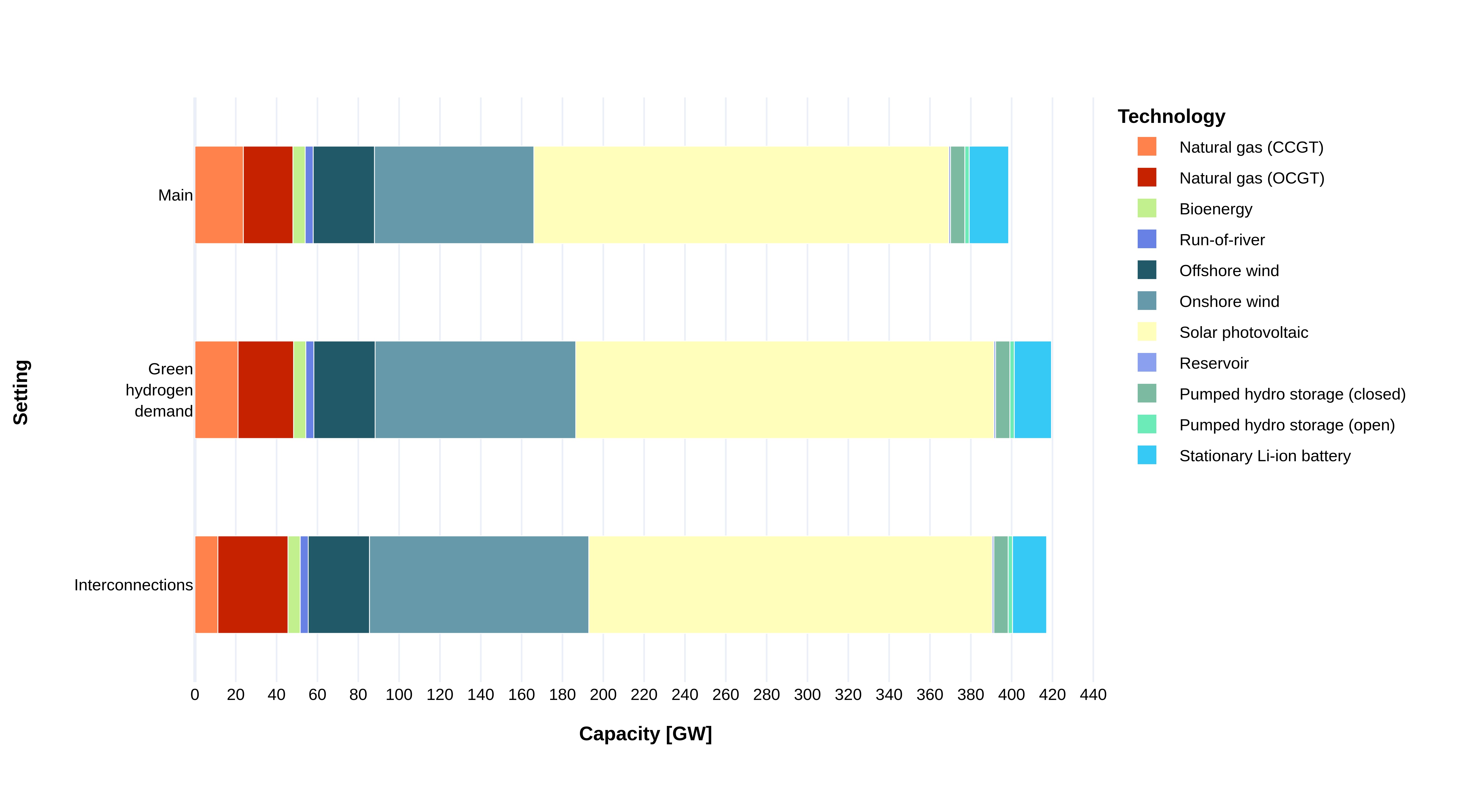}
    \caption{
    \textbf{Capacity mix by technology for the reference in the main setting and two alternative settings.} 
    The two alternative settings include one setting with an industrial green hydrogen demand (``Green hydrogen demand'') and one with interconnected neighboring power sectors (``Interconnections''). In all settings, the reference refers to a system without BEVs.
    }
    \label{fig:capacity_reference}
\end{figure}

\begin{figure}[!ht]
    \centering
    \includegraphics[width=0.95\linewidth]{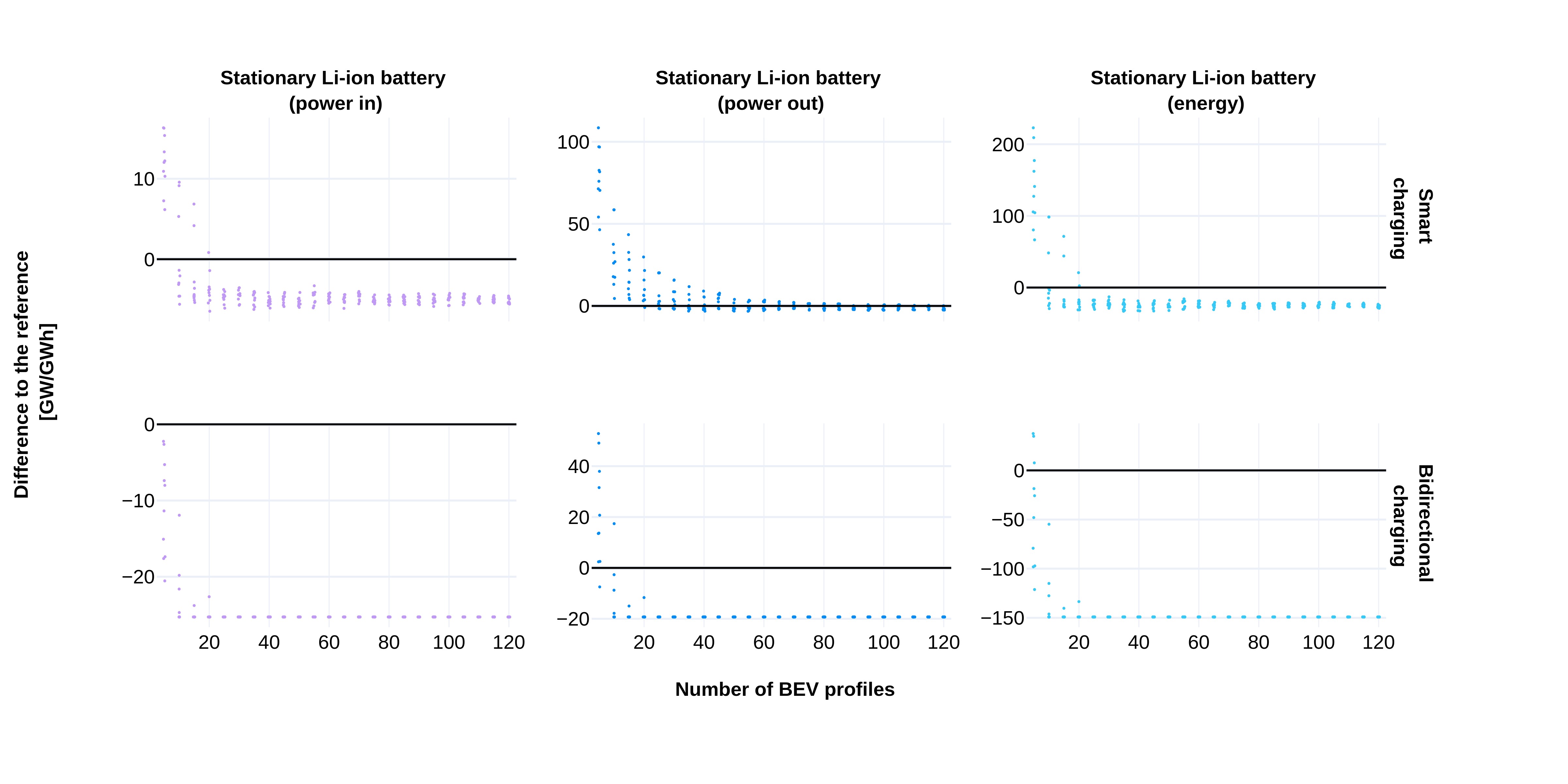}
    \caption{
    \textbf{Change in optimal power and energy capacity for stationary Li-ion battery by charging strategy [Main setting].}
    Changes are expressed relative to the reference with no BEVs. For $x$ a given number of BEV profiles, each dot refers to the optimal solution of a model including a randomly drawn sample of $x$ BEV profiles.
    }
    \label{fig:capacity_battery}
\end{figure}

\clearpage

\begin{figure}[!ht]
    \centering
    \includegraphics[width=0.95\linewidth]{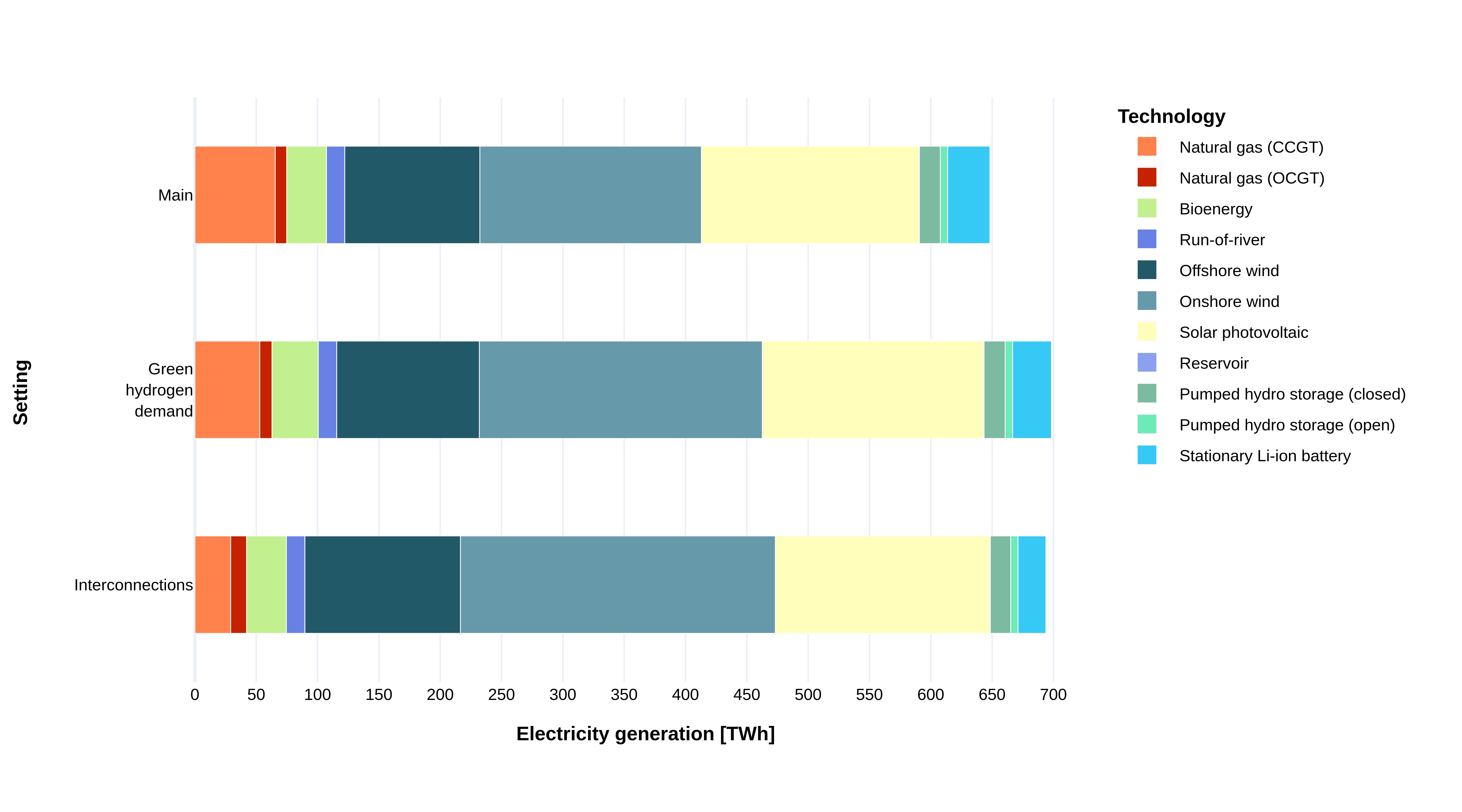}
    \caption{
    \textbf{Electricity generation mix by technology for the reference in the main setting and two alternative settings.} 
    The two alternative settings include one setting with an industrial green hydrogen demand (``Green hydrogen demand'') and one with interconnected neighboring power sectors (``Interconnections''). In all settings, the reference refers to a system without BEVs.
    }
    \label{fig:dispatch_reference}
\end{figure}

\begin{figure}[!ht]
    \centering
    \includegraphics[width=1\linewidth]{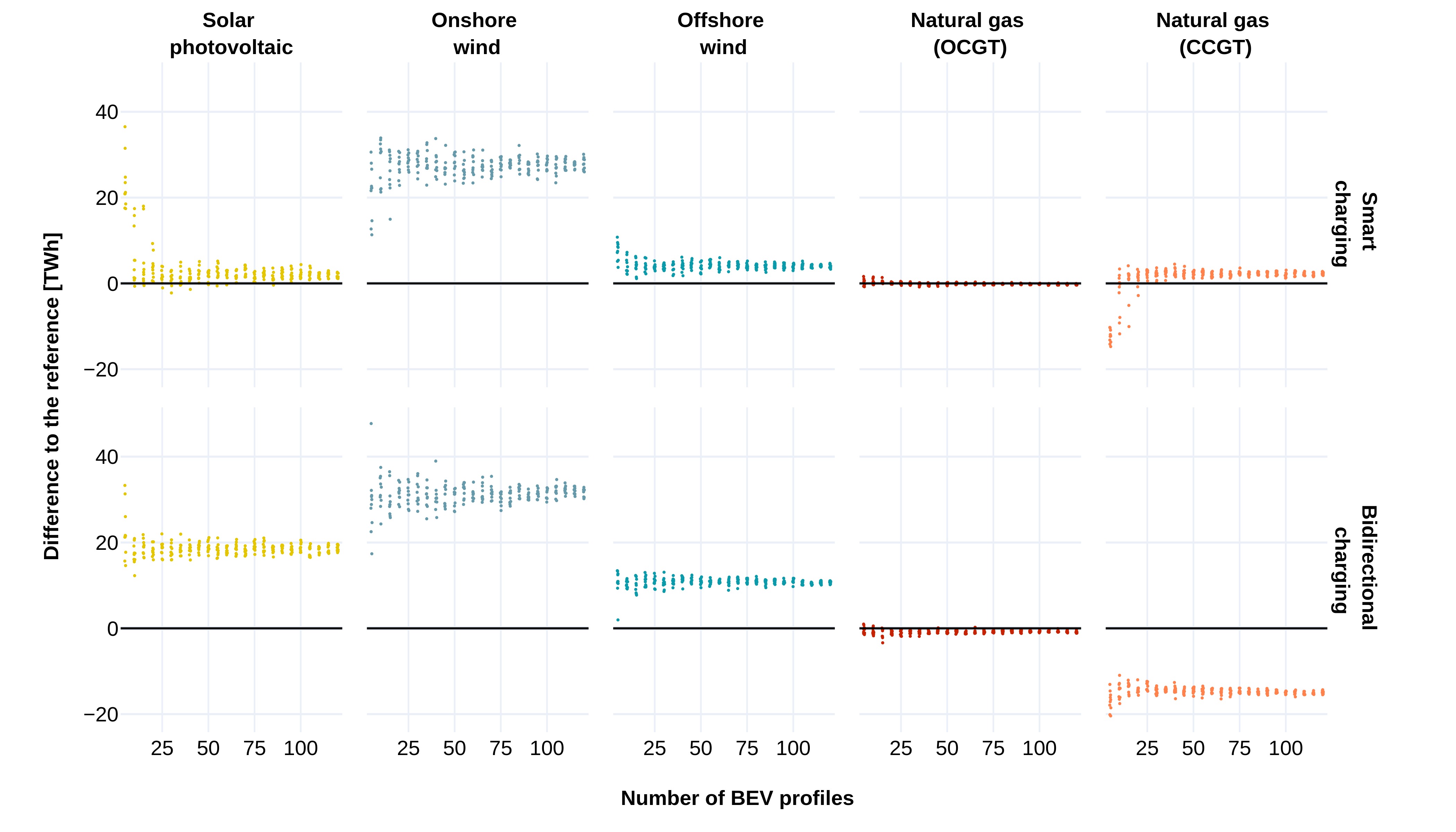}
    \caption{
    \textbf{Change in optimal aggregate electricity generation for selected technologies by charging strategy [Main setting].}
    Changes are expressed relative to the reference with no BEVs. For $x$ a given number of BEV profiles, each dot refers to the optimal solution of a model including a randomly drawn sample of $x$ BEV profiles.
    }
    \label{fig:dispatch_generation}
\end{figure}

\begin{figure}[!ht]
    \centering
    \includegraphics[width=1\linewidth]{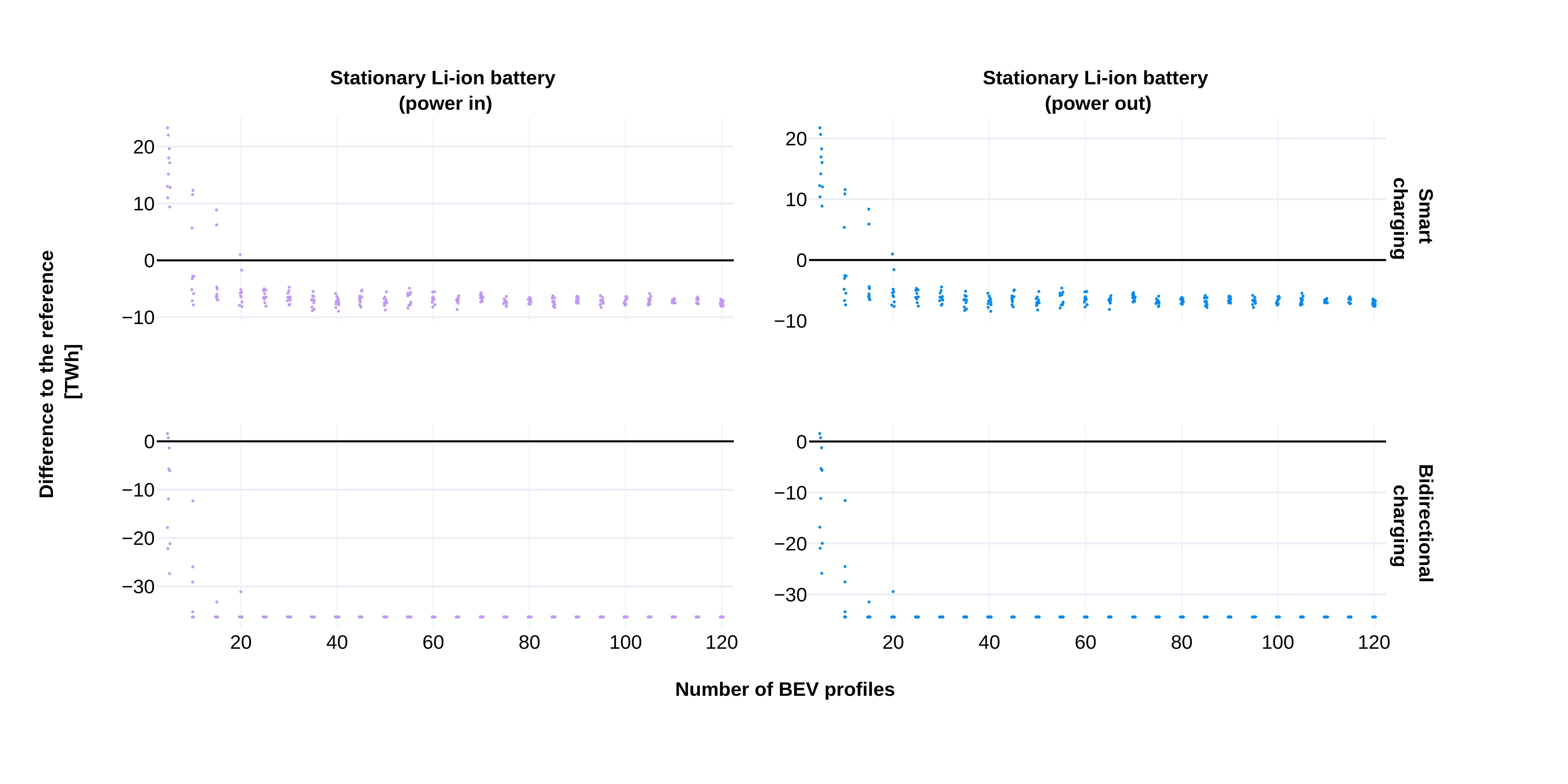}
    \caption{
    \textbf{Change in optimal aggregate charging and discharging for stationary Li-ion battery by charging strategy [Main setting].}
    Changes are expressed relative to the reference with no BEVs. For $x$ a given number of BEV profiles, each dot refers to the optimal solution of a model including a randomly drawn sample of $x$ BEV profiles.
    }
    \label{fig:dispatch_battery}
\end{figure}

\clearpage

\begin{figure}[!ht]
    \centering
    \includegraphics[width=.8\linewidth]{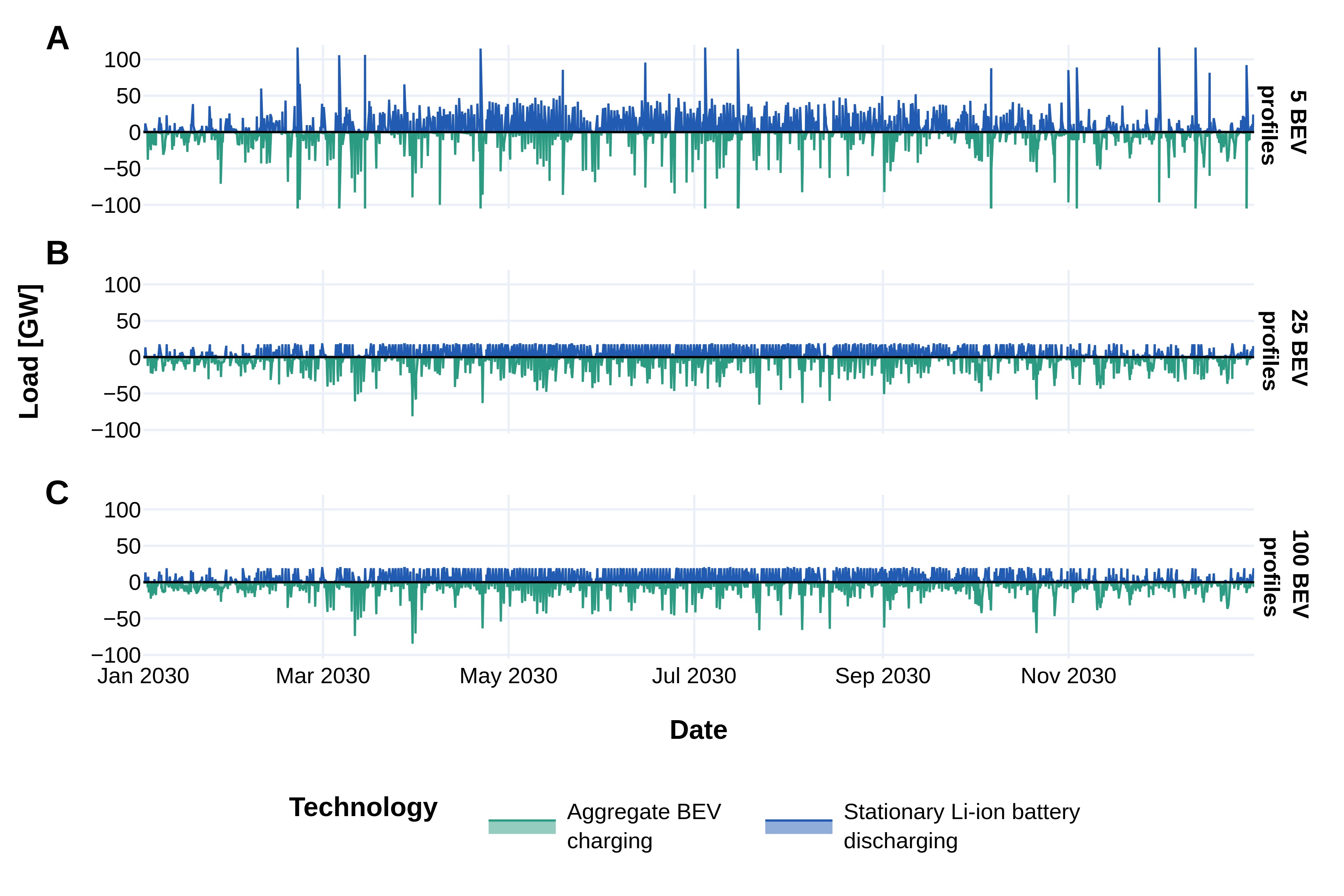}
    \caption{
    \textbf{Aggregate BEV charging and stationary Li-ion battery discharging for selected numbers of profiles over the full year [Main setting].} The charging strategy considered is smart charging. Selected numbers of profiles are: 5 profiles (\textbf{A}), 25 profiles (\textbf{B}) and 100 profiles (\textbf{C}).
    }
    \label{fig:timeseries_year}
\end{figure}

\begin{figure}[!ht]
    \centering
    \includegraphics[width=0.8\linewidth]{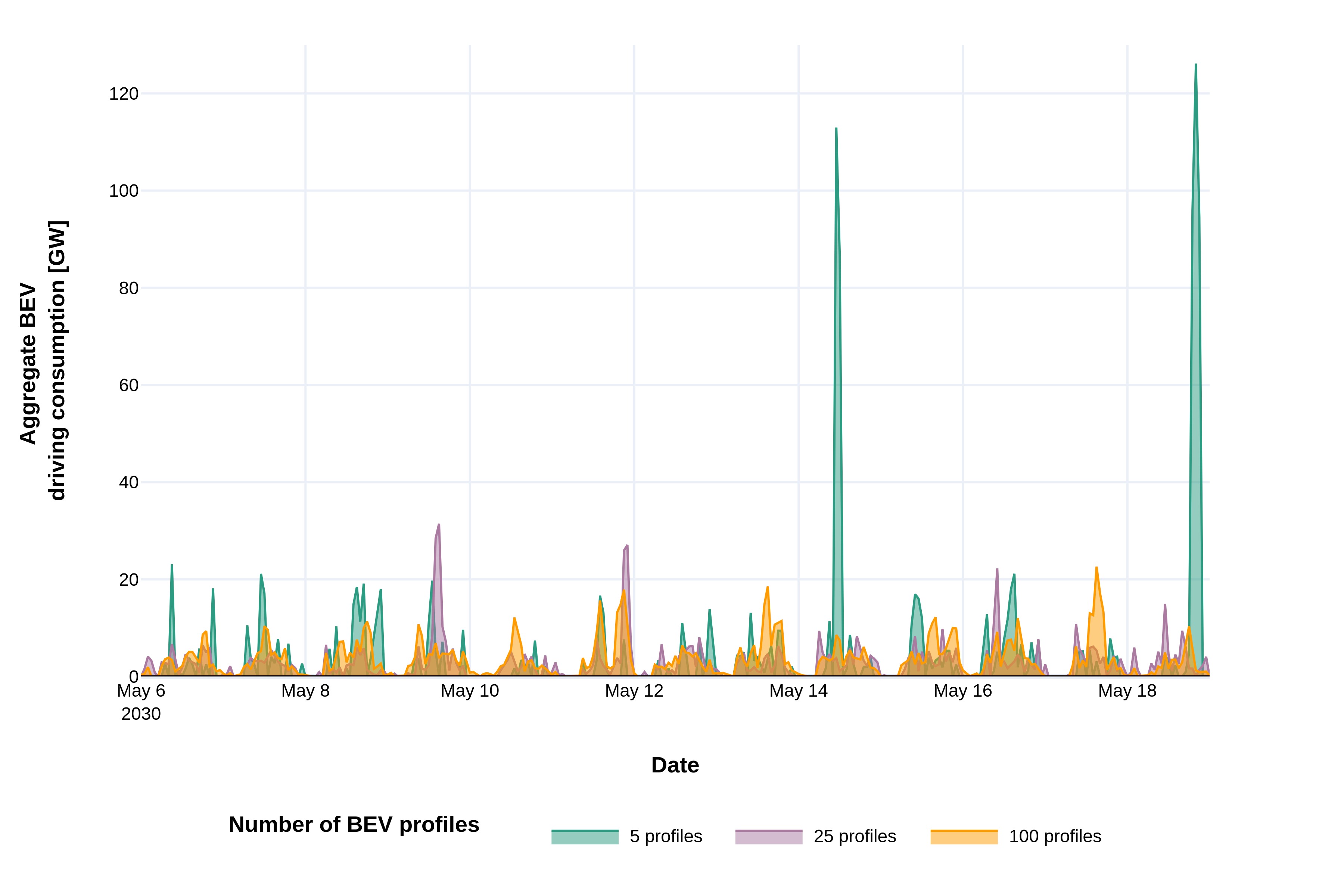}
    \caption{
    \textbf{Aggregate BEV driving consumption for selected number of profiles and a subset of spring days.}
    }
    \label{fig:timeseries_evload}
\end{figure}

\clearpage

\begin{figure}[!ht]
    \centering
    \includegraphics[width=0.95\linewidth]{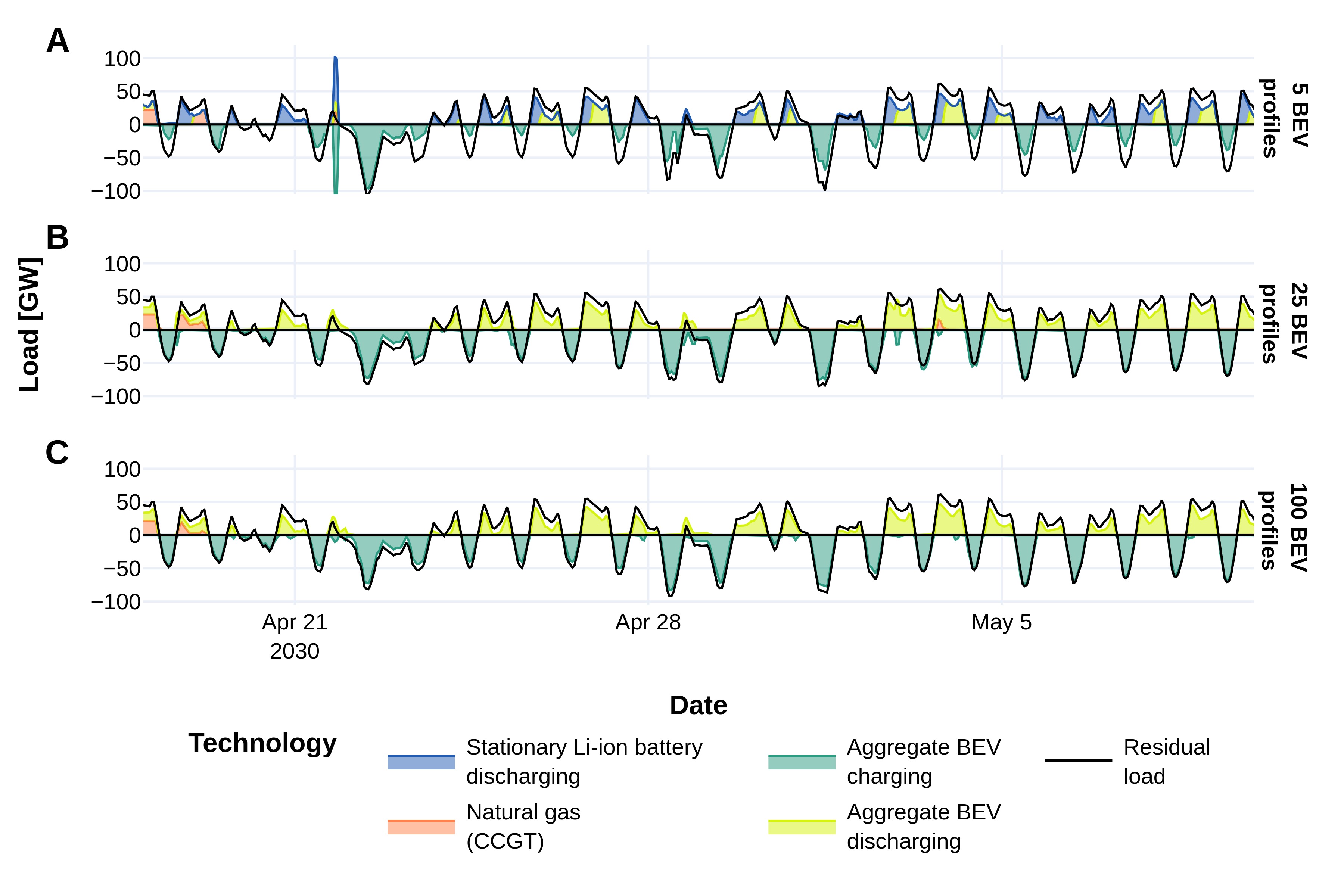}
    \caption{
    \textbf{Optimal dispatch of selected technologies and residual load for a selection of number of BEV profiles and a subset of spring days [Main setting, bidirectional charging].}
    The selection of number of profiles displayed corresponds to 5 profiles (\textbf{A}), 25 profiles (\textbf{B}) and 100 profiles (\textbf{C}). The residual load computation considers vRES generation after curtailment and does not include the load for BEV charging nor BEV discharging.
    }
    \label{fig:timeseries_dispatch_v2g}
\end{figure}

\begin{figure}[!ht]
    \centering
    \includegraphics[width=1\linewidth]{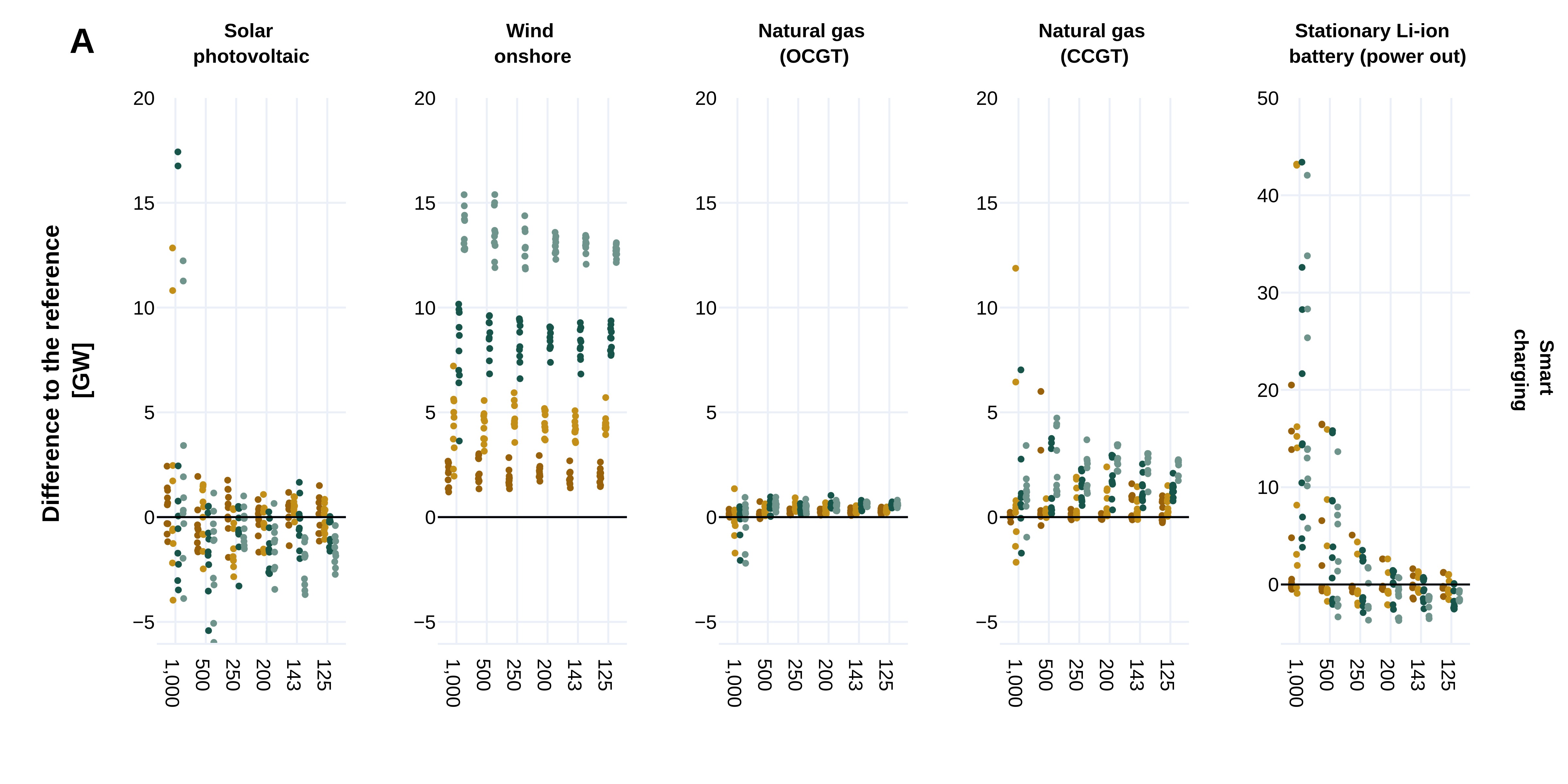}
    \includegraphics[width=1\linewidth]{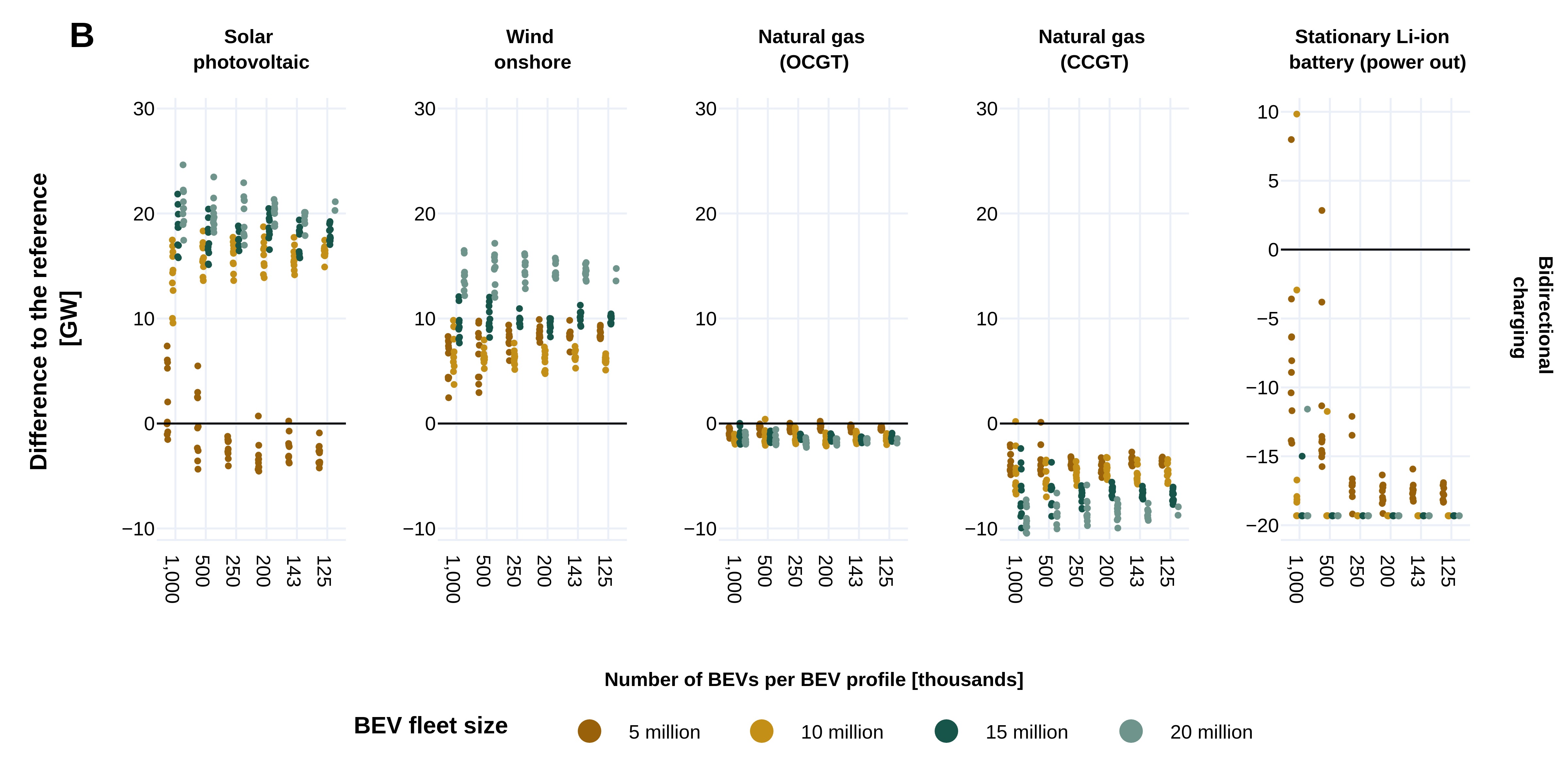}
    \caption{
    \textbf{Change in optimal capacity for selected technologies by BEV fleet size and charging strategy [Main setting].}
    Changes are expressed relative to the reference with no BEVs. For $x$ a given number of BEV profiles, each dot refers to the optimal solution of a model including a randomly drawn sample of $x$ BEV profiles. }
    \label{fig:capacity_fleetsizes}
\end{figure}

\begin{figure}[!ht]
    \centering
    \includegraphics[width=1\linewidth]{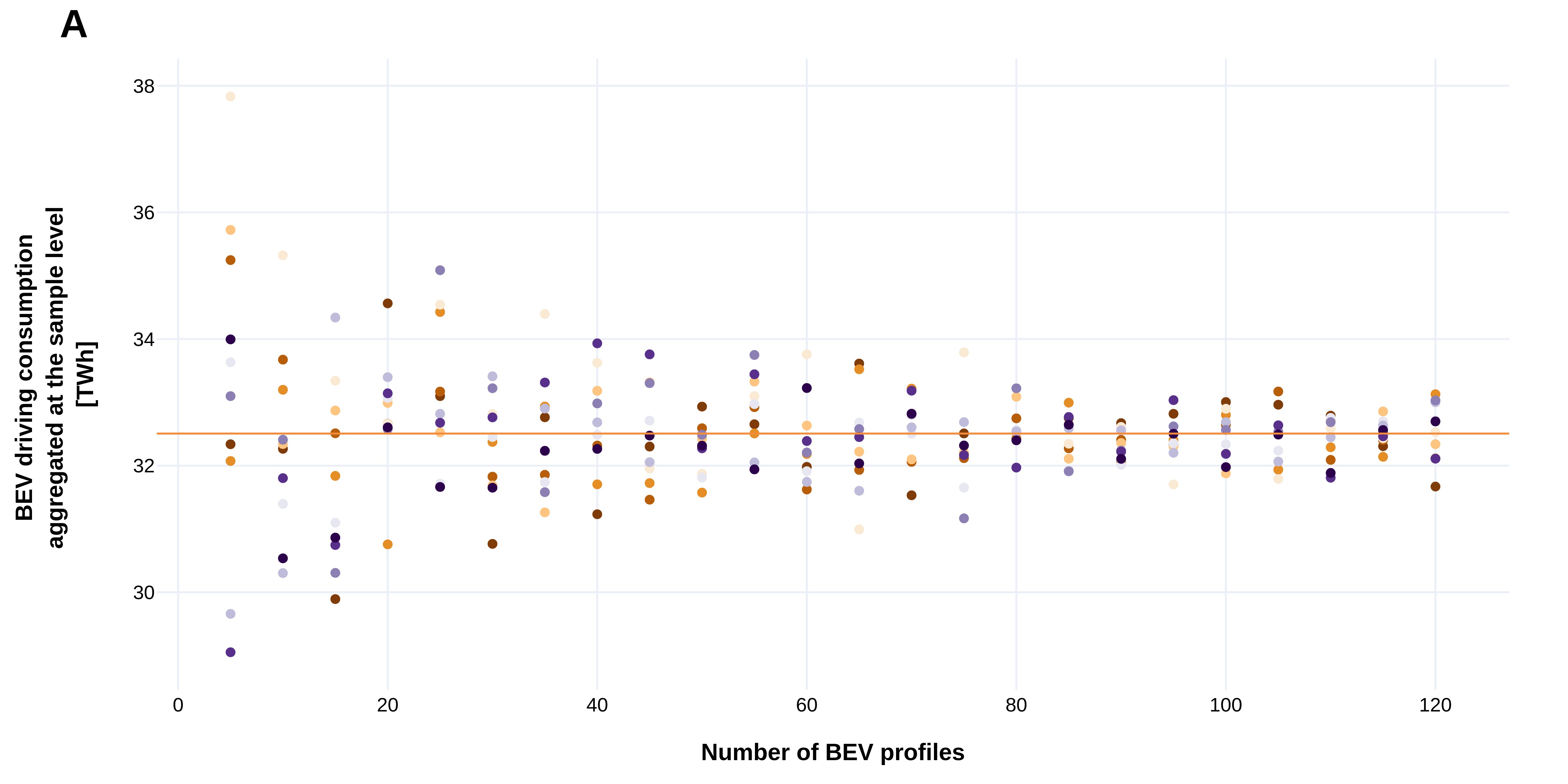}
    \includegraphics[width=1\linewidth]{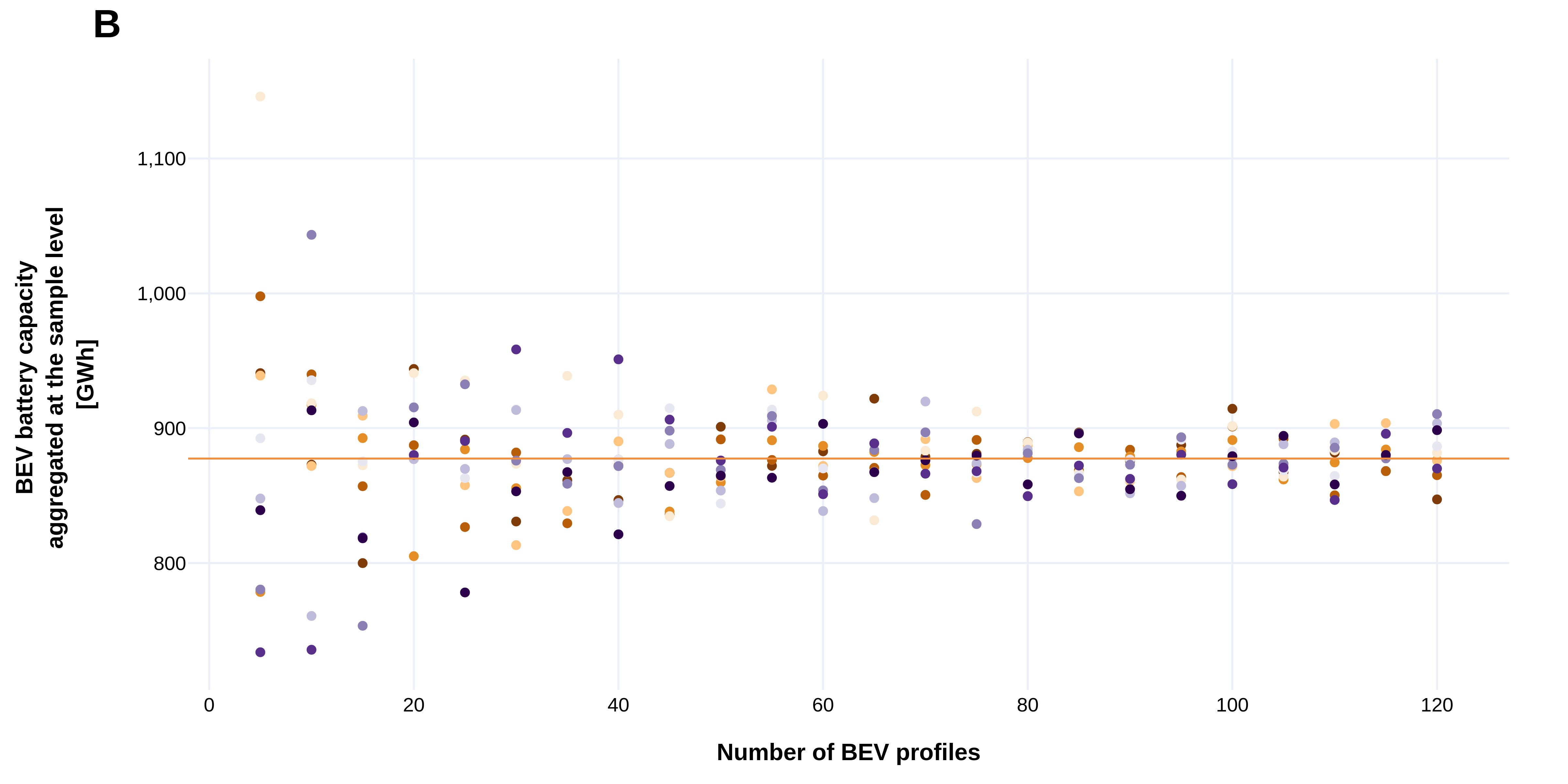}
    \caption{
    \textbf{Aggregate BEV characteristics by sample ID and number of BEV profiles.} 
    Aggregate BEV characteristics refer to the BEV electricity driving consumption aggregated over all scaled BEV profiles within each sample (\textbf{A}) and, similarly, to the BEV battery capacity aggregated over all scaled BEV profiles within each sample (\textbf{B}). For each aggregate characteristic, the solid orange line depicts the average computed across all samples. Colors refer to the sample ID. Samples with similar colors in panels A and B are identical.
    }
    \label{fig:bev_characteristics_absolute_consumption_capacity}
\end{figure}

\clearpage
\subsection{Input data}

\begin{table}[!ht]
\centering
\resizebox{\textwidth}{!}{%
\begin{tabular}{@{}
>{\columncolor[HTML]{FFFFFF}}c 
>{\columncolor[HTML]{FFFFFF}}c 
>{\columncolor[HTML]{FFFFFF}}c 
>{\columncolor[HTML]{FFFFFF}}c 
>{\columncolor[HTML]{FFFFFF}}c 
>{\columncolor[HTML]{FFFFFF}}c 
>{\columncolor[HTML]{FFFFFF}}c 
>{\columncolor[HTML]{FFFFFF}}c 
>{\columncolor[HTML]{FFFFFF}}c 
>{\columncolor[HTML]{FFFFFF}}c 
>{\columncolor[HTML]{FFFFFF}}c 
>{\columncolor[HTML]{FFFFFF}}c @{}}
\toprule
\textbf{} &
  \textbf{Austria} &
  \textbf{Belgium} &
  \textbf{Switzerland} &
  \textbf{Czech Republic} &
  \textbf{Germany} &
  \textbf{Denmark} &
  \textbf{France} &
  \textbf{Italy} &
  \textbf{Luxembourg} &
  \textbf{Netherlands} &
  \textbf{Poland} \\ \midrule
\multicolumn{1}{l|}{\cellcolor[HTML]{FFFFFF}\textbf{Austria}} &
  \textbf{-} &
  - &
  1200 &
  900 &
  7400 &
  - &
  - &
  500 &
  - &
  - &
  - \\
\multicolumn{1}{l|}{\cellcolor[HTML]{FFFFFF}\textbf{Belgium}} &
  - &
  \textbf{-} &
  - &
  - &
  1000 &
  - &
  5300 &
  - &
  180 &
  4400 &
  - \\
\multicolumn{1}{l|}{\cellcolor[HTML]{FFFFFF}\textbf{Switzerland}} &
  1200 &
  - &
  \textbf{-} &
  - &
  6600 &
  - &
  4700 &
  1700 &
  - &
  - &
  - \\
\multicolumn{1}{l|}{\cellcolor[HTML]{FFFFFF}\textbf{Czech Republic}} &
  900 &
  - &
  - &
  \textbf{-} &
  2100 &
  - &
  - &
  - &
  - &
  - &
  800 \\
\multicolumn{1}{l|}{\cellcolor[HTML]{FFFFFF}\textbf{Germany}} &
  7400 &
  1000 &
  6600 &
  2100 &
  \textbf{-} &
  44485 &
  6000 &
  - &
  2300 &
  5000 &
  3000 \\
\multicolumn{1}{l|}{\cellcolor[HTML]{FFFFFF}\textbf{Denmark}} &
  - &
  - &
  - &
  - &
  4485 &
  \textbf{-} &
  - &
  - &
  - &
  2000 &
  - \\
\multicolumn{1}{l|}{\cellcolor[HTML]{FFFFFF}\textbf{France}} &
  - &
  5300 &
  4700 &
  - &
  6000 &
  - &
  \textbf{-} &
  2180 &
  - &
  - &
  - \\
\multicolumn{1}{l|}{\cellcolor[HTML]{FFFFFF}\textbf{Italy}} &
  - &
  - &
  1700 &
  - &
  - &
  - &
  2180 &
  \textbf{-} &
  - &
  - &
  - \\
\multicolumn{1}{l|}{\cellcolor[HTML]{FFFFFF}\textbf{Luxembourg}} &
  - &
  180 &
  - &
  - &
  2300 &
  - &
  - &
  - &
  \textbf{-} &
  - &
  - \\
\multicolumn{1}{l|}{\cellcolor[HTML]{FFFFFF}\textbf{Netherlands}} &
  - &
  4400 &
  - &
  - &
  5000 &
  2000 &
  - &
  - &
  - &
  \textbf{-} &
  - \\
\multicolumn{1}{l|}{\cellcolor[HTML]{FFFFFF}\textbf{Poland}} &
  - &
  - &
  - &
  800 &
  3000 &
  - &
  - &
  - &
  - &
  - &
  \textbf{-} \\ \bottomrule
\end{tabular}%
}
\caption{\textbf{Net Transfer Capacities (NTC) in MW.}}
\label{tab:parameters_ntc}
\end{table}
\begin{table}[!ht]
\centering
\resizebox{\textwidth}{!}{%
\begin{tabular}{@{}
>{\columncolor[HTML]{FFFFFF}}c 
>{\columncolor[HTML]{FFFFFF}}l 
>{\columncolor[HTML]{FFFFFF}}c 
>{\columncolor[HTML]{FFFFFF}}c 
>{\columncolor[HTML]{FFFFFF}}c 
>{\columncolor[HTML]{FFFFFF}}c 
>{\columncolor[HTML]{FFFFFF}}c 
>{\columncolor[HTML]{FFFFFF}}c 
>{\columncolor[HTML]{FFFFFF}}c 
>{\columncolor[HTML]{FFFFFF}}c 
>{\columncolor[HTML]{FFFFFF}}c 
>{\columncolor[HTML]{FFFFFF}}c 
>{\columncolor[HTML]{FFFFFF}}c @{}}
\toprule
\multicolumn{1}{l}{\cellcolor[HTML]{FFFFFF}} &
  \textbf{} &
  \textbf{Germany} &
  \textbf{Austria} &
  \textbf{Belgium} &
  \textbf{Switzerland} &
  \textbf{Czech Republic} &
  \textbf{Denmark} &
  \textbf{France} &
  \textbf{Italy} &
  \textbf{Luxembourg} &
  \textbf{Netherlands} &
  \textbf{Poland} \\ \midrule
\cellcolor[HTML]{FFFFFF} &
  \textit{Lignite} &
  0-9.3 &
  0.0 &
  0.0 &
  0.0 &
  0-3.9 &
  0.0 &
  0.0 &
  0.0 &
  0.0 &
  0.0 &
  0-6.3 \\
\cellcolor[HTML]{FFFFFF} &
  \textit{Hard coal} &
  0-9.8 &
  0.0 &
  0-0.6 &
  0.0 &
  0-0.4 &
  0-0.8 &
  0.0 &
  0.0 &
  0.0 &
  0.0 &
  0-9.9 \\
\cellcolor[HTML]{FFFFFF} &
  \textit{Other fossil fuels} &
  0-4.1 &
  0-0.9 &
  0-1.3 &
  0-0.9 &
  0-1.2 &
  0-0.2 &
  0-1.9 &
  0-6.0 &
  0-0.03 &
  0-3.8 &
  0-6.8 \\
\cellcolor[HTML]{FFFFFF} &
  \textit{Oil} &
  0-1.2 &
  0-0.2 &
  0.0 &
  0.0 &
  0-0.01 &
  0.0 &
  0 &
  0.0 &
  0.0 &
  0.0 &
  0.0 \\
\cellcolor[HTML]{FFFFFF} &
  \textit{Natural gas (CCGT)} &
  0-$\infty$ &
  0-2.8 &
  0-7.6 &
  0.0 &
  0-1.3 &
  0.0 &
  0-6.5 &
  0-3.9 &
  0.0 &
  0-8.6 &
  0-5.0 \\
\cellcolor[HTML]{FFFFFF} &
  \textit{Natural gas (OCGT)} &
  0-$\infty$ &
  0-0.6 &
  0-1.1 &
  0.0 &
  0.0 &
  0.0 &
  0-0.9 &
  0-5.4 &
  0.0 &
  0-0.6 &
  0.0 \\
\cellcolor[HTML]{FFFFFF} &
  \textit{Nuclear} &
  0.0 &
  0.0 &
  0.0 &
  1.2 &
  4.0 &
  0.0 &
  58.2 &
  0.0 &
  0.0 &
  0.5 &
  0.0 \\ \cmidrule(l){2-13} 
\cellcolor[HTML]{FFFFFF} &
  \textit{Bioenergy} &
  6.0 &
  0.6 &
  0.2 &
  1.2 &
  1.1 &
  0.7 &
  2.6 &
  4.9 &
  0.05 &
  0.5 &
  1.4 \\
\cellcolor[HTML]{FFFFFF} &
  \textit{Run-of-river} &
  3.9 &
  6.4 &
  0.2 &
  4.2 &
  0.4 &
  0.0 &
  13.6 &
  7.0 &
  0.04 &
  0.04 &
  0.4 \\
\cellcolor[HTML]{FFFFFF} &
  \textit{Offshore wind} &
  30.0 &
  0.0 &
  4.3 &
  0.0 &
  0.0 &
  4.8 &
  3.0 &
  0.6 &
  0.0 &
  6.7 &
  0.9 \\
\cellcolor[HTML]{FFFFFF} &
  \textit{Onshore wind} &
  0-$\infty$ &
  10.0 &
  5.9 &
  1.3 &
  3.0 &
  5.5 &
  44.1 &
  19.0 &
  0-0.3 &
  8.3 &
  11.3 \\
\multirow{-12}{*}{\cellcolor[HTML]{FFFFFF}\textbf{Generation technologies}} &
  \textit{Solar photovoltaic} &
  0-$\infty$ &
  15.0 &
  13.9 &
  11.0 &
  10.5 &
  4.7 &
  42.6 &
  49.3 &
  0.3 &
  15.5 &
  12.2 \\ \midrule
\cellcolor[HTML]{FFFFFF} &
  \textit{Reservoir} &
   &
   &
   &
   &
   &
   &
   &
   &
   &
   &
   \\
\cellcolor[HTML]{FFFFFF} &
  \multicolumn{1}{l}{\cellcolor[HTML]{FFFFFF}...power out} &
  0.8 &
  2.8 &
  0.0 &
  8.5 &
  0.2 &
  0.0 &
  9.8 &
  8.8 &
  0.0 &
  0.0 &
  0.4 \\
\cellcolor[HTML]{FFFFFF} &
  \multicolumn{1}{l}{\cellcolor[HTML]{FFFFFF}...energy {[}TWh{]}} &
  0.237 &
  0.769 &
  0.0 &
  7.912 &
  0.002 &
  0.0 &
  10.0 &
  5.568 &
  0.0 &
  0.0 &
  0.001 \\
\cellcolor[HTML]{FFFFFF} &
  \textit{Pumped hydro storage (closed)} &
   &
   &
   &
   &
   &
   &
   &
   &
   &
   &
   \\
\cellcolor[HTML]{FFFFFF} &
  \multicolumn{1}{l}{\cellcolor[HTML]{FFFFFF}...power in/out} &
  7.17/7.01 &
  0.45/0.45 &
  1.23/1.31 &
  1.90/1.90 &
  0.66/0.69 &
  0.0/0.0 &
  1.95/1.95 &
  4.17/4.17 &
  0.0/0.0 &
  0.0/0.0 &
  1.49/1.32 \\
\cellcolor[HTML]{FFFFFF} &
  \multicolumn{1}{l}{\cellcolor[HTML]{FFFFFF}...energy {[}TWh{]}} &
  0.392 &
  0.004 &
  0.006 &
  0.056 &
  0.004 &
  0.0 &
  0.010 &
  0.061 &
  0.0 &
  0.0 &
  0.006 \\
\cellcolor[HTML]{FFFFFF} &
  \textit{Pumped hydro storage (open)} &
   &
   &
   &
   &
   &
   &
   &
   &
   &
   &
   \\
\cellcolor[HTML]{FFFFFF} &
  \multicolumn{1}{l}{\cellcolor[HTML]{FFFFFF}...power in/out} &
  1.86/2.14 &
  5.33/5.61 &
  0.0/0.0 &
  1.89/2.46 &
  0.60/0.65 &
  0.0/0.0 &
  1.85/1.85 &
  2.22/3.62 &
  0.0/0.0 &
  0.0/0.0 &
  0.17/0.22 \\
\cellcolor[HTML]{FFFFFF} &
  \multicolumn{1}{l}{\cellcolor[HTML]{FFFFFF}...energy {[}TWh{]}} &
  0.471 &
  1.747 &
  0.0 &
  1.194 &
  0.003 &
  0.0 &
  0.090 &
  0.290 &
  0.0 &
  0.0 &
  0.001 \\
\cellcolor[HTML]{FFFFFF} &
  \textit{Stationary Li-ion battery} &
   &
   &
   &
   &
   &
   &
   &
   &
   &
   &
   \\
\cellcolor[HTML]{FFFFFF} &
  \multicolumn{1}{l}{\cellcolor[HTML]{FFFFFF}...power in/out} &
  0-$\infty$/0-$\infty$ &
  0-$\infty$/0-$\infty$ &
  0-$\infty$/0-$\infty$ &
  0-$\infty$/0-$\infty$ &
  0-$\infty$/0-$\infty$ &
  0-$\infty$/0-$\infty$ &
  0-$\infty$/0-$\infty$ &
  0-$\infty$/0-$\infty$ &
  0-$\infty$/0-$\infty$ &
  0-$\infty$/0-$\infty$ &
  0-$\infty$/0-$\infty$ \\
\cellcolor[HTML]{FFFFFF} &
  \multicolumn{1}{l}{\cellcolor[HTML]{FFFFFF}...energy {[}TWh{]}} &
  0-$\infty$ &
  0-$\infty$ &
  0-$\infty$ &
  0-$\infty$ &
  0-$\infty$ &
  0-$\infty$ &
  0-$\infty$ &
  0-$\infty$ &
  0-$\infty$ &
  0-$\infty$ &
  0-$\infty$ \\ \cmidrule(l){2-13} 
\cellcolor[HTML]{FFFFFF} &
  \textit{PEM electrolyzer} &
   &
   &
   &
   &
   &
   &
   &
   &
   &
   &
   \\
\cellcolor[HTML]{FFFFFF} &
  \multicolumn{1}{l}{\cellcolor[HTML]{FFFFFF}...power} &
  0-$\infty$ &
  0-$\infty$ &
  0-$\infty$ &
  0-$\infty$ &
  0-$\infty$ &
  0-$\infty$ &
  0-$\infty$ &
  0-$\infty$ &
  0-$\infty$ &
  0-$\infty$ &
  0-$\infty$ \\
\cellcolor[HTML]{FFFFFF} &
  \textit{Hydrogen cavern} &
   &
   &
   &
   &
   &
   &
   &
   &
   &
   &
   \\
\cellcolor[HTML]{FFFFFF} &
  \multicolumn{1}{l}{\cellcolor[HTML]{FFFFFF}...energy {[}TWh{]}} &
  35731.0 &
  0.0 &
  0.0 &
  0.0 &
  0.0 &
  7698.3 &
  511.0 &
  0.0 &
  0.0 &
  10420.9 &
  7256.9 \\
\cellcolor[HTML]{FFFFFF} &
  \textit{Hydrogen (OCGT)} &
   &
   &
   &
   &
   &
   &
   &
   &
   &
   &
   \\
\multirow{-18}{*}{\cellcolor[HTML]{FFFFFF}\textbf{Storage technologies}} &
  \multicolumn{1}{l}{\cellcolor[HTML]{FFFFFF}...power} &
  0-$\infty$ &
  0 &
  0 &
  0 &
  0 &
  0-$\infty$ &
  0-$\infty$ &
  0 &
  0 &
  0-$\infty$ &
  0-$\infty$ \\ \bottomrule
\end{tabular}%
}
\caption{\textbf{Capacity bounds by country and technology.} Based on ENTSO-E. ``TYNDP 2018. Project Sheets''. Tech. rep. 2018. Unless otherwise specified, capacity bounds are expressed in gigawatts (GW). For a given technology and country, two numbers separated by an hyphen refer to endogenous capacity investments within this range.}
\label{tab:parameters_capacitybounds}
\end{table}
\begin{table}[!ht]
\centering
\resizebox{\textwidth}{!}{%
\begin{tabular}{@{}lrrrrrr@{}}
\toprule
\rowcolor[HTML]{FFFFFF} 
\multicolumn{1}{c}{\cellcolor[HTML]{FFFFFF}} &
  \multicolumn{3}{c}{\cellcolor[HTML]{FFFFFF}\textbf{Costs}} &
  \multicolumn{1}{c}{\cellcolor[HTML]{FFFFFF}} &
  \multicolumn{1}{c}{\cellcolor[HTML]{FFFFFF}} &
  \multicolumn{1}{c}{\cellcolor[HTML]{FFFFFF}} \\ \cmidrule(lr){2-4}
\rowcolor[HTML]{FFFFFF} 
\multicolumn{1}{c}{\cellcolor[HTML]{FFFFFF}} &
  \multicolumn{1}{c}{\cellcolor[HTML]{FFFFFF}Investment} &
  \multicolumn{1}{c}{\cellcolor[HTML]{FFFFFF}Fixed O\&M} &
  \multicolumn{1}{c}{\cellcolor[HTML]{FFFFFF}Variable \& fuel} &
  \multicolumn{1}{c}{\multirow{-2}{*}{\cellcolor[HTML]{FFFFFF}\textbf{Efficiency}}} &
  \multicolumn{1}{c}{\multirow{-2}{*}{\cellcolor[HTML]{FFFFFF}\textbf{Lifetime}}} &
  \multicolumn{1}{c}{\multirow{-2}{*}{\cellcolor[HTML]{FFFFFF}\textbf{Carbon content}}} \\
\rowcolor[HTML]{FFFFFF} 
\multicolumn{1}{c}{\multirow{-3}{*}{\cellcolor[HTML]{FFFFFF}\textbf{Technology}}} &
  \multicolumn{1}{c}{\cellcolor[HTML]{FFFFFF}{[}\euro/kW{]}} &
  \multicolumn{1}{c}{\cellcolor[HTML]{FFFFFF}{[}\euro/kW/a{]}} &
  \multicolumn{1}{c}{\cellcolor[HTML]{FFFFFF}{[}\euro/MWh{]}} &
  \multicolumn{1}{c}{\cellcolor[HTML]{FFFFFF}} &
  \multicolumn{1}{c}{\cellcolor[HTML]{FFFFFF}{[}years{]}} &
  \multicolumn{1}{c}{\cellcolor[HTML]{FFFFFF}{[}t/MWh{]}} \\ \midrule
\cellcolor[HTML]{FFFFFF}Lignite &
  \cellcolor[HTML]{FFFFFF}1,500.00 &
  30.00 &
  \cellcolor[HTML]{FFFFFF}13.68 &
  0.390 &
  \cellcolor[HTML]{FFFFFF}40 &
  \cellcolor[HTML]{FFFFFF}0.399 \\
\rowcolor[HTML]{FFFFFF} 
Hard coal          & 4,575.69 & 59.98  & 13.15 & 0.405 & 40 & 0.337 \\
\rowcolor[HTML]{FFFFFF} 
Other fossil fuels        & 1,500.00 & 15.00  & 15.10 & 0.350 & 25 & 0.350 \\
\rowcolor[HTML]{FFFFFF} 
Oil                & 361.55   & 8.98   & 49.54 & 0.350 & 25 & 0.257 \\
\rowcolor[HTML]{FFFFFF} 
Natural gas (CCGT) & 882.60   & 29.56  & 21.06 & 0.580 & 25 & 0.201 \\
\rowcolor[HTML]{FFFFFF} 
Natural gas (OCGT) & 467.88   & 8.24   & 26.06 & 0.410 & 25 & 0.201 \\
\rowcolor[HTML]{FFFFFF} 
Nuclear            & 8,594.14 & 109.15 & 5.24  & 0.337 & 40 & 0.000 \\
\rowcolor[HTML]{FFFFFF} 
Bioenergy          & 2,209.00 & 100.00 & 13.65 & 0.468 & 30 & 0.000 \\
\rowcolor[HTML]{FFFFFF} 
Run-of-river       & 3,000.00 & 60.00  & 0.00  & 1.000 & 50 & 0.000 \\ 
\rowcolor[HTML]{FFFFFF} 
Offshore wind      & 1,800.00 & 39.00  & 4.00  & 1.000 & 30 & 0.000 \\
\rowcolor[HTML]{FFFFFF} 
Onshore wind       & 1,146.64 & 16.66  & 1.98  & 1.000 & 30 & 0.000 \\
\rowcolor[HTML]{FFFFFF} 
Solar photovoltaic       & 380.00   & 9.50   & 0.00  & 1.000 & 40 & 0.000 \\ \bottomrule
\end{tabular}%
}
\caption{\textbf{Cost and technology assumptions for electricity generation technologies.} The assumed interest rate is 0.04 and the assumed carbon price is 130 euros per tCO$_2$eq.}
\label{tab:parameters_generation}
\end{table}
\begin{table}[!ht]
\centering
\resizebox{\textwidth}{!}{%
\begin{tabular}{@{}
>{\columncolor[HTML]{FFFFFF}}l 
>{\columncolor[HTML]{FFFFFF}}r 
>{\columncolor[HTML]{FFFFFF}}r 
>{\columncolor[HTML]{FFFFFF}}r 
>{\columncolor[HTML]{FFFFFF}}r 
>{\columncolor[HTML]{FFFFFF}}r 
>{\columncolor[HTML]{FFFFFF}}r 
>{\columncolor[HTML]{FFFFFF}}r 
>{\columncolor[HTML]{FFFFFF}}r 
>{\columncolor[HTML]{FFFFFF}}r 
>{\columncolor[HTML]{FFFFFF}}r @{}}
\toprule
\multicolumn{1}{c}{\cellcolor[HTML]{FFFFFF}} &
  \multicolumn{3}{c}{\cellcolor[HTML]{FFFFFF}\textbf{Investment costs}} &
  \multicolumn{3}{c}{\cellcolor[HTML]{FFFFFF}\textbf{Fixed costs}} &
  \multicolumn{1}{c}{\cellcolor[HTML]{FFFFFF}\textbf{Variable costs}} &
  \multicolumn{2}{c}{\cellcolor[HTML]{FFFFFF}\textbf{Efficiency}} &
  \multicolumn{1}{c}{\cellcolor[HTML]{FFFFFF}} \\ \cmidrule(lr){2-10}
\multicolumn{1}{c}{\cellcolor[HTML]{FFFFFF}} &
  \multicolumn{1}{c}{\cellcolor[HTML]{FFFFFF}Energy} &
  \multicolumn{1}{c}{\cellcolor[HTML]{FFFFFF}Charging} &
  \multicolumn{1}{c}{\cellcolor[HTML]{FFFFFF}Discharging} &
  \multicolumn{1}{c}{\cellcolor[HTML]{FFFFFF}Energy} &
  \multicolumn{1}{c}{\cellcolor[HTML]{FFFFFF}Charging} &
  \multicolumn{1}{c}{\cellcolor[HTML]{FFFFFF}Discharging} &
  \multicolumn{1}{c}{\cellcolor[HTML]{FFFFFF}} &
  \multicolumn{1}{c}{\cellcolor[HTML]{FFFFFF}Charging} &
  \multicolumn{1}{c}{\cellcolor[HTML]{FFFFFF}Discharging} &
  \multicolumn{1}{c}{\multirow{-2}{*}{\cellcolor[HTML]{FFFFFF}\textbf{Lifetime}}} \\
\multicolumn{1}{c}{\multirow{-3}{*}{\cellcolor[HTML]{FFFFFF}\textbf{Technology}}} &
  \multicolumn{1}{c}{\cellcolor[HTML]{FFFFFF}{[}EUR/kWh{]}} &
  \multicolumn{1}{c}{\cellcolor[HTML]{FFFFFF}{[}EUR/kW{]}} &
  \multicolumn{1}{c}{\cellcolor[HTML]{FFFFFF}{[}EUR/kW{]}} &
  \multicolumn{1}{c}{\cellcolor[HTML]{FFFFFF}{[}EUR/kWh/a{]}} &
  \multicolumn{1}{c}{\cellcolor[HTML]{FFFFFF}{[}EUR/kW/a{]}} &
  \multicolumn{1}{c}{\cellcolor[HTML]{FFFFFF}{[}EUR/kW/a{]}} &
  \multicolumn{1}{c}{\cellcolor[HTML]{FFFFFF}{[}EUR/MW{]}} &
  \multicolumn{1}{c}{\cellcolor[HTML]{FFFFFF}} &
  \multicolumn{1}{c}{\cellcolor[HTML]{FFFFFF}} &
  \multicolumn{1}{c}{\cellcolor[HTML]{FFFFFF}{[}years{]}} \\ \midrule
Reservoir                     & 56.03  & 0.00   & 685.81 & 0.00  & 0.00  & 30.00 & 0.10 & -     & 0.950 & 60   \\
Pumped hydro storage (closed)  & 56.03  & 685.81 & 685.81 & 13.65 & 0.00  & 0.00  & 0.50 & 0.894 & 0.894 & 60  \\
Pumped hydro storage (open)    & 56.03  & 685.81 & 685.81 & 13.65 & 0.00  & 0.00  & 0.10 & 0.894 & 0.894 & 60  \\
Stationary Li-ion battery      & 151.00 & 85.08  & 85.08  & 0.00  & 0.29  & 0.29  & 0.96 & 0.985 & 0.975 & 25  \\
Long-duration storage         &        &        &        &       &       &       &      &       &       &           \\
… PEM electrolyzer            & -      & 650.00 & -      & -     & 13.00 & -     & -    & 0.585 & -     & 25  \\
… Hydrogen cavern compressor  & -      & 80.17  & -      & -     & 3.21  & -     & 0.00 & 0.995 & -     & 15   \\
… Hydrogen cavern storage      & 2.13   & -      & -      & 0.003 & -     & -     & -    & -     & -     & 100  \\
… Hydrogen (OCGT)            & -      & -      & 538.07 & -     & -     & 8.24  & 5.00 & -     & 0.410 & 25   \\ \bottomrule
\end{tabular}%
}
\caption{\textbf{Cost and technology assumptions for electricity storage technologies.} The assumed interest rate is 0.04.}
\label{tab:parameters_storage}
\end{table}
\begin{table}[!ht]
\centering
\begin{tabular}{@{}
>{\columncolor[HTML]{FFFFFF}}r 
>{\columncolor[HTML]{FFFFFF}}c @{}}
\toprule
\multicolumn{1}{c}{\cellcolor[HTML]{FFFFFF}\textbf{Country}} & \textbf{Electricity demand {[}TWh{]}} \\ \midrule
\textit{Luxembourg}     & 8.8   \\
\textit{Denmark}        & 52.6  \\
\textit{Switzerland}    & 64.0  \\
\textit{Czech Republic} & 74.1  \\
\textit{Austria}        & 82.2  \\
\textit{Belgium}        & 95.2  \\
\textit{Netherlands}    & 139.9 \\
\textit{Poland}         & 180.9 \\
\textit{Italy}          & 330.6 \\
\textit{France}         & 478.0 \\
\textit{Germany}        & 583.1 \\ \bottomrule
\end{tabular}
\caption{\textbf{Yearly electricity demand of non-BEV consumers by country in TWh.} Data are taken from the ENTSO-E Pan-European Climate Database (PECD 2021.3).}
\label{tab:parameters_load}
\end{table}

\end{document}